\documentclass[preprint,pre,aps,showpacs,showkeys,amsmath]{revtex4}
\usepackage{graphicx}

\begin{document}

\title{Dynamical Responses to External Stimuli for Both Cases of Excitatory and Inhibitory Synchronization in A Complex Neuronal Network}

\author{Sang-Yoon Kim}
\email{sykim@icn.re.kr}
\author{Woochang Lim}
\email{wclim@icn.re.kr}
\affiliation{Institute for Computational Neuroscience and Department of Science Education, Daegu National University of Education, Daegu 42411, Korea}

\begin{abstract}
For studying how dynamical responses to external stimuli depend on the synaptic-coupling type, we consider two types of excitatory and inhibitory synchronization (i.e., synchronization via synaptic excitation and inhibition) in complex small-world networks of excitatory regular spiking (RS) pyramidal neurons and inhibitory fast spiking (FS) interneurons. For both cases of excitatory and inhibitory synchronization, effects of synaptic couplings on dynamical responses to external time-periodic stimuli $S(t)$ (applied to a fraction of neurons) are investigated by varying the driving amplitude $A$ of $S(t)$. Stimulated neurons are phase-locked to external stimuli for both cases of excitatory and inhibitory couplings. On the other hand, the stimulation effect on non-stimulated neurons depends on the type of synaptic coupling. The external stimulus $S(t)$ makes a constructive effect on excitatory non-stimulated RS neurons (i.e., it causes external phase lockings in the non-stimulated sub-population), while $S(t)$ makes a destructive effect on inhibitory non-stimulated FS interneurons (i.e., it breaks up original inhibitory synchronization in the non-stimulated sub-population). As results of these different effects of $S(t)$, the type and degree of dynamical response (e.g., synchronization enhancement or suppression), characterized by the dynamical response factor $D_f$ (given by the ratio of synchronization degree in the presence and absence of stimulus), are found to vary in a distinctly different way, depending on the synaptic-coupling type. Furthermore, we also measure the matching degree between the dynamics of the two sub-populations of stimulated and non-stimulated neurons in terms of a ``cross-correlation'' measure $M_c$. With increasing $A$, based on $M_c$, we discuss the cross-correlations between the two sub-populations, affecting the dynamical responses to $S(t)$.
\end{abstract}

\pacs{87.19.lm, 87.19.lc}
\keywords{Excitatory synchronization, Inhibitory synchronization, External time-periodic stimulus, Dynamical response factor, Cross-correlation measure, Synchronization enhancement, Synchronization suppression}

\maketitle

\section{Introduction}
\label{sec:INT}
Recently, much attention has been paid to brain rhythms in health and diseases \cite{Buz}. These brain rhythms emerge via synchronization between individual firings in neural circuits. This kind of neural synchronization may be used for efficient sensory and cognitive processing such as sensory perception, multisensory integration, selective attention, and memory formation \cite{Wang1,Wang2,Gray}, and it is also correlated with pathological rhythms associated with neural diseases (e.g., epileptic seizures and tremors in the Parkinson's disease) \cite{TW}. The brain receives natural sensory stimulation, and experimental electrical or magnetic stimulation in the neural system is used for analyzing the dynamical interactions between different brain areas. Responses to these external stimuli can provide crucial information about its dynamical properties. For example, the effects of periodic stimuli on rhythmic biological activity were experimentally studied by applying rhythmic visual stimulus \cite{VS} and periodic auditory stimulation \cite{AS}. Hence, it is of great importance to investigate how an external stimulus affects the neural synchronization in the brain. Techniques for controlling population synchronization have been proposed, which enables us to suppress or to enhance it. For examples, one technique is the external time-periodic stimulation \cite{TPS1,TPS2,TPS3,TPTD}, and the other one is the time-delayed feedback in the mean field \cite{TPTD,TDF1,TDF2,TDF3}. Synchronization suppression may be effective in suppressing pathological brain rhythms, while synchronization enhancement might be useful for the cases of failures of cardiac or neural pacemakers. Particularly, deep brain stimulation techniques have been used to suppress pathological rhythms in patients with neural diseases such as Parkinson's disease, essential tremor, and epilepsy \cite{DBS1,DBS2,DBS3}. For this technique, micro-electrodes are implanted in deep brain regions of patients, and then time-periodic electric signal or time-delayed feedback signals are injected for suppression of abnormal rhythms.

Most of previous theoretical and computational works on control of population synchronization were focused on the case of excitatory-type couplings \cite{TPS1,TPS2,TPS3,TPTD,TDF1,TDF2,TDF3}. To study how dynamical responses to external stimuli depend on the synaptic type (excitatory or inhibitory), we consider two types of excitatory and inhibitory full synchronization [i.e., full synchronization (where all neurons fire in each global cycle of population rhythm) via excitatory and inhibitory synaptic interactions] in complex small-world networks of excitatory regular spiking (RS) pyramidal neurons and inhibitory fast spiking (FS) interneurons. We apply external time-periodic stimuli $S(t)$ $[=A \sin (\omega_d t)]$ to a fraction of neurons for both cases of excitatory and inhibitory synchronization, and investigate their dynamical responses to $S(t)$ by changing the driving amplitude $A$ for a fixed driving angular frequency $\omega_d$. For describing collective behaviors in the whole population, we use an instantaneous whole-population spike rate (IWPSR) $R_w(t)$ which may be obtained from the raster plot of spikes where population synchronization may be well seen \cite{Kim1}. For the case of synchronization, $R_w(t)$ shows an oscillatory behaviors, while it becomes nearly stationary in the case of desynchronization. We characterize dynamical responses to $S(t)$ in terms of a dynamical response factor $D_f$ (given by the square root of the ratio of the variance of $R_w(t)$ in the presence and absence of stimulus). If $D_f$ is larger than 1, then synchronization enhancement occurs; otherwise (i.e., $D_f <1$), synchronization suppression takes place. For both cases of excitatory and inhibitory couplings, stimulated neurons are phase-locked to external stimuli $S(t)$. In contrast, the stimulation effect on non-stimulated neurons varies depending on the synaptic-coupling type. For the excitatory case, non-stimulated RS neurons are also phase-locked to external stimulus $S(t)$ thanks to a constructive effect of $S(t)$ (resulting from phase-attractive synaptic excitation). On the other hand, in the inhibitory case the original full synchronization in the non-stimulated sub-population breaks up gradually with increasing $A$ due to a destructive effect of $S(t)$ (coming from strong synaptic inhibition), and then a new type of sparse synchronization (where only some fraction of neurons fire in each global cycle of population rhythm) appears. As results of these different effects of $S(t)$, the type and degree of dynamical response (characterized by $D_f$) vary differently, depending on the type of synaptic interaction. For further analysis of dynamical response, we also decompose the whole population into two sub-populations of the stimulated and the non-stimulated neurons. Then, two instantaneous sub-population spike rates (ISPSRs) $R_s^{(1)}(t)$ and $R_s^{(2)}(t)$
[the superscript 1 (2) corresponds to the stimulated (non-stimulated) case] may be used to show collective behaviors in the two sub-populations of stimulated and non-stimulated neurons, respectively, and the matching degree between the dynamics of the stimulated and the non-stimulated sub-populations is measured in terms of a ``cross-correlation'' measure $M_c$ between $R_s^{(1)}(t)$ and $R_s^{(2)}(t)$. $M_c$ also varies with $A$ in a distinctly different way, depending on the synaptic-coupling type, because of different effects of $S(t)$. Based on the cross-correlations between the two sub-populations (characterized by $M_c$), we also discuss the dynamical responses to $S(t)$.

This paper is organized as follows. In Sec.~\ref{sec:SWN}, we describe complex small-world networks of excitatory RS pyramidal neurons and inhibitory FS interneurons, and the governing equations for the population dynamics are given. Then, in Sec.~\ref{sec:DR} we investigate the effects of synaptic couplings on dynamical responses to external time-periodic stimuli $S(t)$ for both excitatory and inhibitory cases. Finally, in Sec.~\ref{sec:SUM} a summary is given. Explanations on methods for characterization of synchronization in each of the stimulated and the non-stimulated sub-populations are also made in Appendix \ref{sec:SMSM}.

\section{Small-World Networks of Excitatory RS Pyramidal Neurons and Inhibitory FS Interneurons}
\label{sec:SWN}
We consider two types of directed Watts-Strogatz small-world networks (SWNs) composed of $N$ excitatory RS pyramidal neurons and inhibitory FS interneurons equidistantly placed on a one-dimensional ring of radius $N/ 2 \pi$, respectively. The Watts-Strogatz SWN interpolates between a regular lattice with high clustering (corresponding to the case of $p=0$) and a random graph with short average path length (corresponding to the case of $p=1$) via random uniform rewiring with the probability $p$ \cite{SWN1,SWN2,SWN3}. For $p=0,$ we start with a directed regular ring lattice with $N$ nodes where each node is coupled to its first $M_{syn}$ neighbors ($M_{syn}/2$ on either side) via outward synapses, and rewire each outward connection uniformly at random over the whole ring with the probability $p$ (without self-connections and duplicate connections). This Watts-Strogatz SWN model may be regarded as a cluster-friendly extension of the random network by reconciling the six degrees of separation (small-worldness) \cite{SDS1,SDS2} with the circle of friends (clustering). As elements in our neural networks, we choose the Izhikevich RS pyramidal neuron and FS interneuron models which are not only biologically plausible, but also computationally efficient \cite{Izhi1,Izhi2,Izhi3,Izhi4}.

The following equations (\ref{eq:PD1})-(\ref{eq:PD8}) govern the population dynamics in the SWNs:
\begin{eqnarray}
C\frac{dv_i}{dt} &=& k (v_i - v_r) (v_i - v_t) - u_i +I_{DC} +D \xi_{i} -I_{syn,i} + S_i(t), \label{eq:PD1} \\
\frac{du_i}{dt} &=& a \{ U(v_i) - u_i \},  \;\;\; i=1, \cdots, N, \label{eq:PD2}
\end{eqnarray}
with the auxiliary after-spike resetting:
\begin{equation}
{\rm if~} v_i \geq v_p,~ {\rm then~} v_i \leftarrow c~ {\rm and~} u_i \leftarrow u_i + d, \label{eq:PD3}
\end{equation}
where
\begin{eqnarray}
U(v) &=& b (v - v_b)~{\rm for~the~ RS~ pyramidal~ neurons}, \label{eq:PD4} \\
     &=& \left\{ \begin{array}{l} 0 {\rm ~for~} v<v_b \\ b(v - v_b)^3 {\rm ~for~} v \ge v_b \end{array} \right. ~{\rm for~ the~ FS~ interneurons},  \label{eq:PD5} \\
I_{syn,i} &=& \frac{J}{d_i^{(in)}} \sum_{j=1 (\ne i)}^N w_{ij} s_j(t) (v_i - V_{syn}), \label{eq:PD6}\\
s_j(t) &=& \sum_{f=1}^{F_j} E(t-t_f^{(j)}-\tau_l);~E(t) = \frac{1}{\tau_d - \tau_r} (e^{-t/\tau_d} - e^{-t/\tau_r}) \Theta(t), \label{eq:PD7} \\
S_i(t) &=& \alpha_i A \sin(\omega_d t). \label{eq:PD8}
\end{eqnarray}
Here, $v_i(t)$ and $u_i(t)$ are the state variables of the $i$th neuron at a time $t$ which represent the membrane potential and the recovery current, respectively. These membrane potential and the recovery variable, $v_i(t)$ and $u_i(t)$, are reset according to Eq.~(\ref{eq:PD3}) when $v_i(t)$ reaches its cutoff value $v_p$. $C$, $v_r$, and $v_t$ in Eq.~(\ref{eq:PD1}) are the membrane capacitance, the resting membrane potential, and the instantaneous threshold potential, respectively. The parameter values used in our computations are listed in Table \ref{tab:Parm}. More details on the Izhikevich RS pyramidal neuron and FS interneuron models, the external stimulus to each Izhikevich neuron, the synaptic currents, the external time-periodic stimulus to sub-populations of randomly-selected neurons, and the numerical method for integration of the governing equations are given in the following subsections.

\subsection{Izhikevich RS Pyramidal Neuron and FS Interneuron Models}
\label{subsec:Izhi}
The Izhikevich model matches neuronal dynamics by tuning the parameters $(k, a, b, c, d)$ instead of matching neuronal electrophysiology, unlike the Hodgkin-Huxley-type conductance-based models \cite{Izhi1,Izhi2,Izhi3,Izhi4}.
The parameters $k$ and $b$ are related to the neuron's rheobase and input resistance, and $a$, $c$, and $d$ are the recovery time constant, the after-spike reset value of $v$, and the after-spike jump value of $u$, respectively. Depending on the values of these parameters, the Izhikevich neuron model may exhibit 20 of the most prominent neuro-computational features of cortical neurons \cite{Izhi1,Izhi2,Izhi3,Izhi4}. Here, we use the parameter values for the RS pyramidal neurons and the FS interneurons in the layer 5 rat visual cortex, which are listed in the 1st and the 2nd items of Table \ref{tab:Parm}  \cite{Izhi3}.

\subsection{External Stimulus to Each Izhikevich Neuron}
\label{subsec:Sti}
Each Izhikevich neuron is stimulated by both a common DC current $I_{DC}$ and an independent Gaussian white noise $\xi_i$ [see the 3rd and the 4th terms in Eq.~(\ref{eq:PD1})]. The Gaussian white noise satisfies $\langle \xi_i(t) \rangle =0$ and $\langle \xi_i(t)~\xi_j(t') \rangle = \delta_{ij}~\delta(t-t')$, where $\langle\cdots\rangle$ denotes the ensemble average. Here, the Gaussian noise $\xi$ may be
regarded as a parametric one which randomly perturbs the strength of the applied current $I_{DC}$, and its intensity is controlled by the parameter $D$.
For $D=0$, the Izhikevich RS pyramidal neurons exhibit the type-I excitability, while the Izhikevich FS interneurons show the type-II excitability \cite{Izhi3}.
For the type-I case, a transition from a resting state to a spiking state occurs as $I_{DC}$ passes a threshold via a saddle-node bifurcation on an invariant circle, and firing begins at arbitrarily low frequency \cite{Izhi3,Ex1,Ex2}. On the other hand, a type-II neuron exhibits a jump from a resting state to a spiking state through a subcritical Hopf bifurcation when passing a threshold by absorbing an unstable limit cycle born via fold limit cycle bifurcation and hence, the firing frequency begins from a non-zero value \cite{Izhi3,Ex1,Ex2}. The values of $I_{DC}$ and $D$ used in this paper are given in the 3rd item of Table \ref{tab:Parm}.

\subsection{Synaptic Currents}
\label{subsec:Syn}
The 5th term in Eq.~(\ref{eq:PD1}) denotes the synaptic couplings of Izhikevich neurons. $I_{syn,i}$ of Eqs.~(\ref{eq:PD6}) represents the synaptic current injected into the $i$th neuron.
The synaptic connectivity is given by the connection weight matrix $W$ (=$\{ w_{ij} \}$) where $w_{ij}=1$ if the neuron $j$ is presynaptic to the neuron $i$; otherwise, $w_{ij}=0$.
Here, the synaptic connection is modeled in terms of the Watts-Strogatz SWN. The in-degree of the $i$th neuron, $d_{i}^{(in)}$ (i.e., the number of synaptic inputs to the neuron $i$) is given by $d_{i}^{(in)} =
\sum_{j=1(\ne i)}^N w_{ij}$. For this case, the average number of synaptic inputs per neuron is given by $M_{syn} = \frac{1}{N} \sum_{i=1}^{N} d_{i}^{(in)}$.

The fraction of open synaptic ion channels at time $t$ is denoted by $s(t)$. The time course of $s_j(t)$ of the $j$th neuron is given by a sum of delayed double-exponential functions $E(t-t_f^{(j)}-\tau_l)$ [see Eq.~(\ref{eq:PD7})], where $\tau_l$ is the synaptic delay, and $t_f^{(j)}$ and $F_j$ are the $f$th spiking time and the total number of spikes of the $j$th neuron (which occur until time $t$), respectively. Here, $E(t)$ [which corresponds to contribution of a presynaptic spike occurring at time $0$ to $s(t)$ in the absence of synaptic delay] is controlled by the two synaptic time constants: synaptic rise time $\tau_r$ and decay time $\tau_d$, and $\Theta(t)$ is the Heaviside step function: $\Theta(t)=1$ for $t \geq 0$ and 0 for $t <0$. The synaptic coupling strength is controlled by the parameter $J$,  and $V_{syn}$ is the synaptic reversal potential.
For both excitatory AMPA synapse and the inhibitory GABAergic synapse (involving the $\rm{GABA_A}$ receptors), the values of $\tau_l$, $\tau_r$, $\tau_d$, and $V_{syn}$ are listed in the 4th item of Table \ref{tab:Parm}
\cite{Synapse}.

\subsection{External Time-Periodic Stimulus to Sub-Populations of Randomly-Selected Neurons}
\label{subsec:ST}
The last term in Eq.~(\ref{eq:PD1}) represents the external time-periodic stimulus to the $i$th neuron, $S_i(t)$, the explicit form of which is given in Eq.~(\ref{eq:PD8}). If stimulus is applied to the $i$th neuron, $\alpha_i=1$; otherwise, $\alpha_i=0.$ (In the absence of external stimulus, $\alpha_i=0$ for all $i$.) The driving angular frequency of the stimulus is $\omega_d$, and its amplitude is $A.$
We apply $S_i(t)$ to sub-groups of randomly-chosen $N_s (=50)$ RS pyramidal neurons and FS interneurons, respectively.

\subsection{Numerical Method for Integration}
\label{subsec:NM}
Numerical integration of stochastic differential Eqs.~(\ref{eq:PD1})-(\ref{eq:PD8}) is done by employing the Heun method \cite{SDE} with the time step $\Delta t=0.01$ ms. For each realization of the stochastic process, we choose  random initial points $[v_i(0),u_i(0)]$ for the $i$th $(i=1,\dots, N)$ RS pyramidal neuron and FS interneuron with uniform probability in the range of $v_i(0) \in (-50,-45)$ and $u_i(0) \in (10,15)$.

\section{Effects of Synaptic-Coupling Type on Dynamical Responses to External Time-Periodic Stimuli}
\label{sec:DR}
In this section, we study the effects of synaptic-coupling type on dynamical responses to external time-periodic stimuli $S(t)$ in the Watts-Strogatz SWN with the average number of synaptic inputs $M_{syn}=50$ and the rewiring probability $p=0.2$. Both the excitatory and the inhibitory cases are investigated by varying the driving amplitude $A$ for a fixed driving angular frequency $\omega_d$.

\subsection {Dynamical Response of Excitatory Synchronization to An External Time-Periodic Stimulus}
\label{subsec:DREx}
\begin{figure}
\includegraphics[width=0.6\columnwidth]{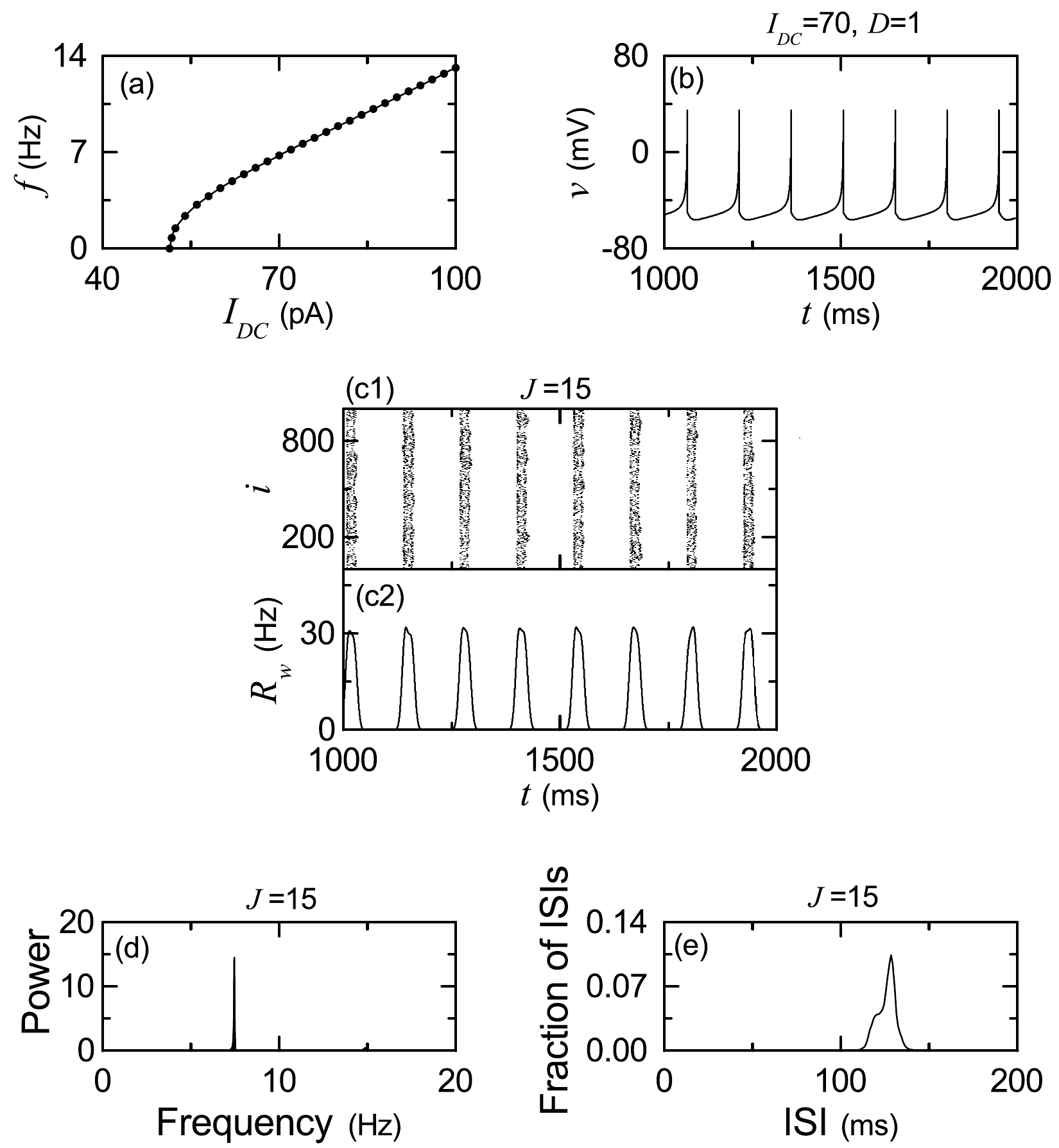}
\caption{Excitatory synchronization in the Watts-Strogatz SWN of $N (=10^3)$ Izhikevich RS pyramidal neurons for $M_{syn}=50$ and $p=0.2$. Single Izhikevich RS pyramidal neuron: (a) plot of the mean firing rate $f$ vs. the external DC current $I_{DC}$ for $D=0$ and (b) time series of the membrane potential $v$ for $I_{DC}=70$ and $D=1$. Coupled Izhikevich RS pyramidal neurons for $I_{DC}=70$, $D=1$, and $J=15$: (c1) raster plot of spikes, (c2) plot of the instantaneous whole-population spike rate (IWPSR) kernel estimate $R_w(t)$ versus $t$, (d) one-sided power spectrum of $\Delta R_w(t) [=R_w(t) - \overline{R_w(t)}]$ (the overbar represents the time average) with mean-squared amplitude normalization, and (e) inter-spike interval (ISI) histogram. The band width of the Gaussian kernel estimate for the IWPSR $R_w(t)$ is 7 ms and the power spectrum is obtained via 30 realizations [$2^{16} (=65536)$ data points are used in each realization]. The ISI histogram is also obtained through 30 realizations ($5 \times 10^7$ ISIs are used in each realization) and the bin size for the histogram is 0.5 ms.
}
\label{fig:RS}
\end{figure}

We consider an excitatory Watts-Strogatz SWN composed of $N (=10^3)$ Izhikevich RS pyramidal neurons. Figure \ref{fig:RS}(a) shows a plot of the firing frequency $f$ versus the external DC current $I_{DC}$ for a single  Izhikevich RS neuron in the absence of noise ($D=0$). This Izhikevich RS neuron exhibits type-I excitability for $I_{DC} > 51$ because its frequency may be arbitrarily small \cite{Izhi3,Ex1,Ex2}. Here, we consider a suprathreshold case of $I_{DC}=70$ in the presence of noise with its intensity $D=1$ for which a time series of the membrane potential $v$ with an oscillating frequency $f \simeq 7.0$ Hz is shown in Fig.~\ref{fig:RS}(b). We set the coupling strength at $J=15$. Spike synchronization is well seen in the raster plot of spikes in Fig.~\ref{fig:RS}(c1). ``Stripes'' (composed of synchronized spikes) appear regularly. All Izhikevich RS neurons fire synchronously in each stripe, and hence full synchronization occurs. For this synchronous case, an oscillating IWPSR (instantaneous whole-population spike rate) $R_w(t)$ appears. To obtain a smooth IWPSR, we employ the kernel density estimation (kernel smoother) \cite{Kernel}. Each spike in the raster plot is convoluted (or blurred) with a kernel function $K_h(t)$ to obtain a smooth estimate of IWPSR $R_w(t)$:
\begin{equation}
R_w(t) = \frac{1}{N} \sum_{i=1}^{N} \sum_{s=1}^{n_i} K_h (t-t_{s}^{(i)}),
\label{eq:IPSR}
\end{equation}
where $t_{s}^{(i)}$ is the $s$th spiking time of the $i$th neuron, $n_i$ is the total number of spikes for the $i$th neuron, and we use a Gaussian kernel function of band width $h$:
\begin{equation}
K_h (t) = \frac{1}{\sqrt{2\pi}h} e^{-t^2 / 2h^2}, ~~~~ -\infty < t < \infty.
\label{eq:KF}
\end{equation}
Figure \ref{fig:RS}(c2) shows a regularly-oscillating IWPSR kernel estimate $R_w(t)$. The population frequency $f_p (\simeq 7.6$ Hz) of $R_w(t)$ may be obtained from the power spectrum of $\Delta R_w(t)$ $[= R_w(t) - \overline{R_w(t)}]$ (the overline represents the time average), which is shown in Fig.~\ref{fig:RS}(d). For analysis of individual spiking behaviors, an inter-spike interval (ISI) histogram is given in Fig.~\ref{fig:RS}(e).
The ensemble-averaged ISI $\langle ISI \rangle$ ($\langle \cdots \rangle$ denotes an ensemble average) is 131.6 ms, and hence the ensemble-averaged mean firing rate (MFR) $\langle f_i \rangle$ of individual neurons ($f_i$ is the MFR of the $i$th neuron and $\langle f_i \rangle$ corresponds to the reciprocal of $\langle ISI \rangle$) is 7.6 Hz. For the case of full synchronization, $f_p = \langle f_i \rangle$, in contrast to the case of sparse synchronization where $f_p$ is larger than $\langle f_i \rangle$ due to stochastic spike skipping of individual neurons \cite{Sparse,SWN-Kim}.

\begin{figure}
\includegraphics[width=0.8\columnwidth]{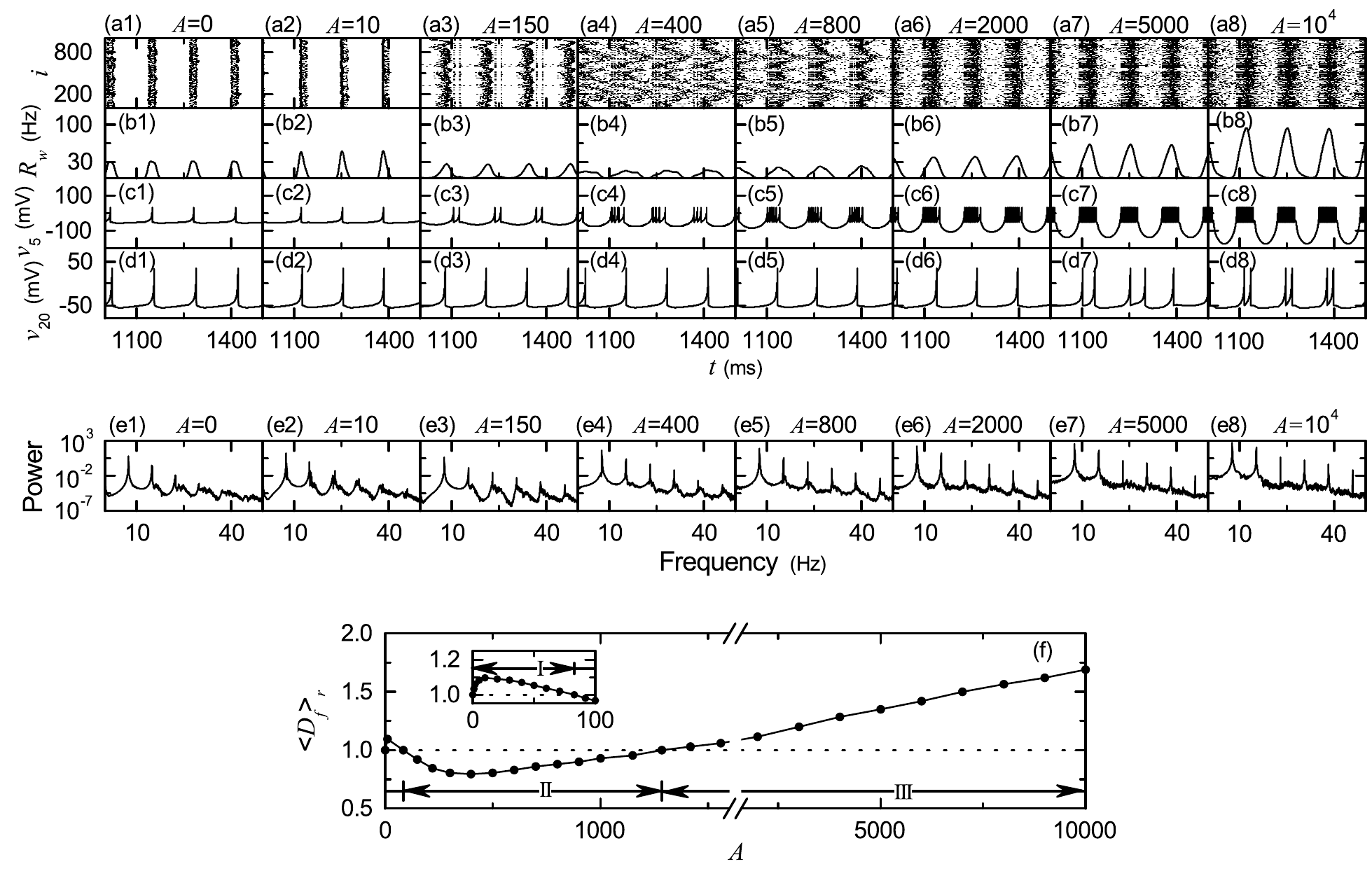}
\caption{Dynamical response when an external time-periodic stimulus $S(t)$ is applied to 50 randomly-chosen Izhikevich RS pyramidal neurons in the case of excitatory synchronization for $J=15$ in Fig.~\ref{fig:RS}. We vary the driving amplitude $A$ for a fixed driving angular frequency $\omega_d$ (=0.048 rad/ms). Raster plots of spikes, instantaneous whole-population spike rate (IWPSR) kernel estimates $R_w(t)$, membrane potentials $v_5(t)$ of the stimulated 5th RS pyramidal neuron, and membrane potentials $v_{20}(t)$ of the non-stimulated 20th RS pyramidal neuron are shown for various values of $A$ in (a1)-(a8), (b1)-(b8), (c1)-(c8), and (d1)-(d8), respectively. One-sided power spectra of $\Delta R_w(t) [=R_w(t) - \overline{R_w(t)}]$ (the overbar represents the time average) with mean-squared amplitude normalization are also given in (e1)-(e8). The band width of the Gaussian kernel estimate for each IWPSR $R_w(t)$ is 7 ms, and each power spectrum is obtained via 30 realizations [$2^{16} (=65536)$ data points are used in each realization]. (f) Plot of dynamical response factor $\langle D_f \rangle_r$ versus $A$, where I, II, and III represent the 1st (synchronization enhancement), the 2nd (synchronization suppression), and the 3rd (synchronization enhancement) stages, respectively. Here, $\langle \cdots \rangle_r$ represents an average over 30 realizations. Averaging time for $D_f$ in each realization is $3 \times 10^4$ ms.
}
\label{fig:DREx1}
\end{figure}

We apply an external time-periodic AC stimulus $S(t)$ to a sub-population of $N_s(=50)$ randomly-selected Izhikevich RS pyramidal neurons by fixing the driving angular frequency as $\omega_d (=2 \pi f_d)$ =0.048 rad/ms ($f_d=\langle f_i \rangle=$ 7.6 Hz), and investigate the dynamical response of the above full synchronization for $J=15$ by varying the driving amplitude $A$. Figures \ref{fig:DREx1}(a1)-\ref{fig:DREx1}(a8) show raster plots of spikes for various values of $A$. Their corresponding IWPSR kernel estimates $R_w(t)$ are shown in Figs.~\ref{fig:DREx1}(b1)-\ref{fig:DREx1}(b8), and the power spectra of $\Delta R_w(t)$ are also given in Figs.~\ref{fig:DREx1}(e1)-\ref{fig:DREx1}(e8). Population synchronization may be well seen in these raster plots of spikes. For a synchronous case, the IWPSR kernel estimates $R_w(t)$ exhibits an oscillating behavior. In addition, times series of individual membrane potentials $v_5(t)$ and $v_{20}(t)$ of the stimulated 5th and the non-stimulated 20th RS neurons are also given in Figs.~\ref{fig:DREx1}(c1)-\ref{fig:DREx1}(c8) and Figs.~\ref{fig:DREx1}(d1)-\ref{fig:DREx1}(d8), respectively. Then, the type and degree of dynamical response may be characterized in terms of a dynamical response factor $D_f$ \cite{TDF1,TDF2}:
\begin{equation}
D_f = \sqrt{\frac {Var(R_w^{(A)})} {Var(R_w^{(0)})} },
\label{eq:DRF}
\end{equation}
where $Var(R_w^{(A)})$ and $Var(R_w^{(0)})$ represent the variances of the IWPSR kernel estimate $R_w(t)$ in the presence and absence of stimulus, respectively. If the dynamical response factor $D_f$ is larger than 1, then synchronization enhancement occurs; otherwise (i.e., $D_f <1$), synchronization suppression takes place. Figure \ref{fig:DREx1}(f) shows a plot of $\langle D_f \rangle_r$ versus $A$; $\langle \cdots \rangle_r$ denotes an average over realizations. Three stages are found to appear. Synchronization enhancement ($\langle D_f \rangle_r >1$), synchronization suppression ($\langle D_f \rangle_r <1$), and synchronization enhancement (i.e., increase in $\langle D_f \rangle_r$ from 1) occur in the 1st (I) stage ($0< A < A^*_{1}$), the 2nd (II) stage ($A^*_{1} < A < A^*_{2}$), and the 3rd (III) stage ($A>A^*_{2}$), respectively; $A^*_{1}\simeq 83$ and $A^*_{2} \simeq 1287$. Examples are given for various values of $A$; 1st stage ($A=10$), 2nd stage ($A=150$, 400 and 800), and 3rd stage $(A=2000,$ 5000, and $10^4$).

\begin{figure}
\includegraphics[width=0.8\columnwidth]{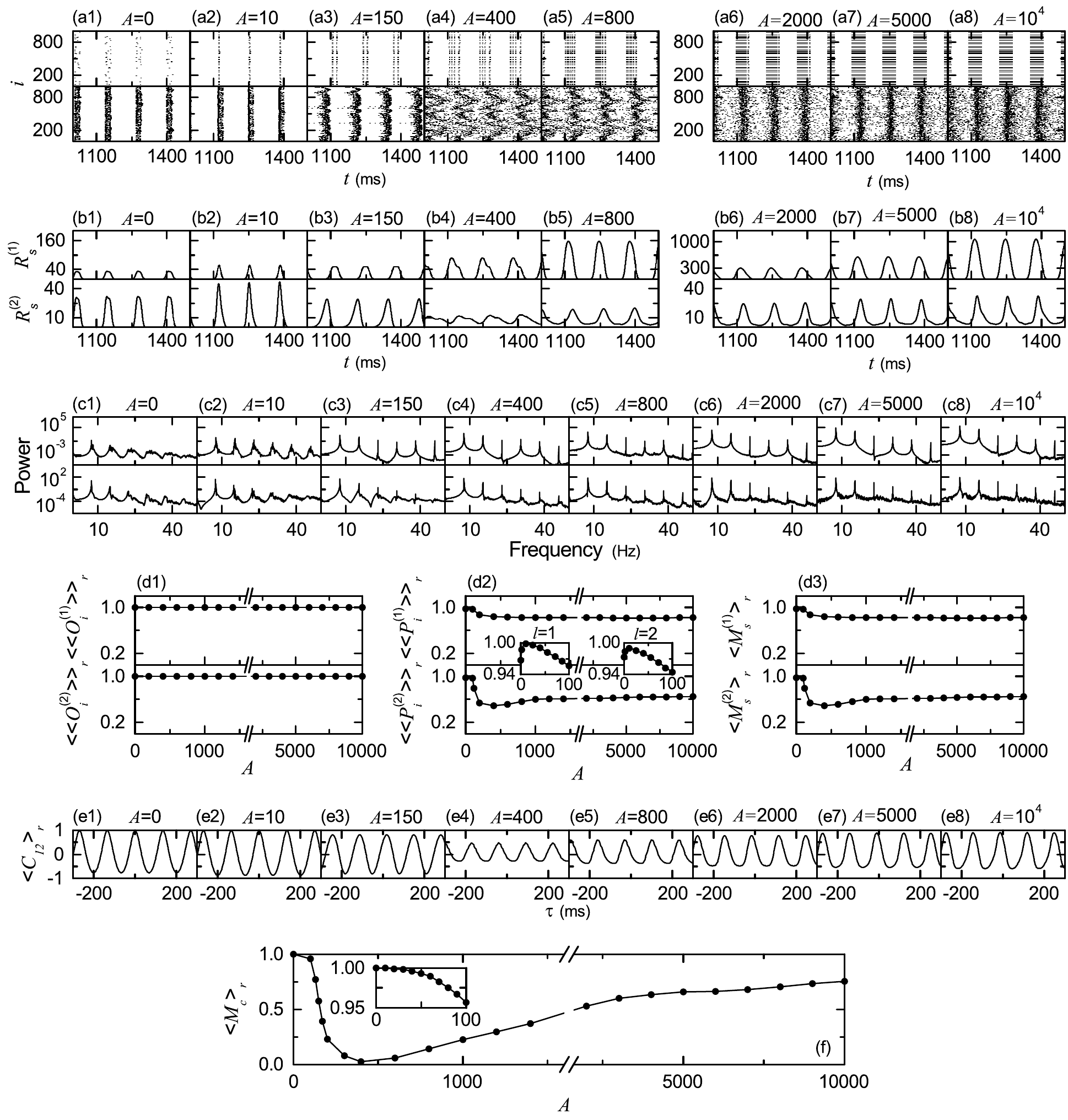}
\linespread{0.86}
\caption{Dynamical responses for the case of Fig.~\ref{fig:DREx1} in both the stimulated and the non-stimulated sub-populations and a cross-correlation measure $\langle M_c \rangle_r$ between the dynamics of the two sub-populations. Raster plot of spikes, instantaneous sub-population spike rates (ISPSRs) $R_s^{(1)}$ and $R_s^{(2)}$ [the superscript 1 (2) corresponds to the stimulated (non-stimulated) case], and one-sided power spectra of $\Delta R_s^{(1)}(t) [=R_s^{(1)}(t) - \overline{R_s^{(1)}(t)}]$ and  $\Delta R_s^{(2)}(t) [=R_s^{(2)}(t) - \overline{R_s^{(2)}(t)}]$ (the overbar represents the time average) with mean-squared amplitude normalization in the stimulated and the non-stimulated sub-populations are shown for various values of $A$ in (a1)-(a8), (b1)-(b8), and (c1)-(c8), respectively: the upper (lower) panels denote those for the stimulated (non-stimulated) case. The band width of the Gaussian kernel estimate for each ISPSR is 7 ms, and each power spectrum is obtained via 30 realizations [$2^{16} (=65536)$ data points are used in each realization]. Plots of the average occupation degree $\langle \langle O_i^{(l)} \rangle \rangle_r$, the average pacing degree $\langle \langle P_i^{(l)} \rangle \rangle_r$, and the statistical-mechanical spiking measure $\langle M_s^{(l)} \rangle_r$ versus $A$ are shown in (d1)-(d3), respectively; $l=1$ (2) corresponds to the stimulated (non-stimulated) case. Here, $\langle O_i^{(l)} \rangle$, $\langle P_i^{(l)} \rangle$, and $M_s^{(l)}$ are obtained by following the $3 \times 10^3$ stripes/bands in the raster plot of spikes in each realization, and $\langle \cdots \rangle_r$ denotes an average over 30 realizations. Plots of cross-correlation functions $\langle C_{12}(\tau) \rangle_r$ versus $\tau$ are also given in (e1)-(e8). The number of data used for the calculation of each temporal cross-correlation function $C_{12}(\tau)$ is $2^{16} (=65536)$ in each realization, and $\langle C_{12}(\tau) \rangle_r$ is obtained through an average over 30 realizations. (f) Plot of the cross-correlation measure $\langle M_c \rangle_r$ versus $A$; $\langle M_c \rangle_r$ is obtained via an average over 30 realizations.
}
\label{fig:DREx2}
\end{figure}

For further analysis of dynamical responses, we decompose the whole population of RS neurons into two sub-populations of the stimulated and the non-stimulated RS neurons. Dynamical responses in these two sub-populations are shown well in Fig.~\ref{fig:DREx2}. Raster plots of spikes, instantaneous sub-population spike rate (ISPSR) kernel estimates $R_s^{(1)}(t)$ and $R_s^{(2)}(t)$ [the superscript 1 (2) corresponds to the stimulated (non-stimulated) case], and power spectra of $\Delta R_s^{(1)}(t)$ and $\Delta R_s^{(2)}(t)$ in the stimulated and the non-stimulated sub-populations are shown in Figs.~\ref{fig:DREx2}(a1)-\ref{fig:DREx2}(a8), Figs.~\ref{fig:DREx2}(b1)-\ref{fig:DREx2}(b8), and Figs.~\ref{fig:DREx2}(c1)-\ref{fig:DREx2}(c8), respectively: the upper (lower) panels in these figures represent those for the stimulated (non-stimulated) case. We also measure the degree of population synchronization in each of the stimulated and the non-stimulated sub-populations by employing a realistic statistical-mechanical spiking measure, which was developed in our recent work \cite{Kim1}. As shown in Figs.~\ref{fig:DREx2}(a1)-\ref{fig:DREx2}(a8), population synchronization may be well visualized in a raster plot of spikes. For a synchronized case, the raster plot is composed of spiking stripes or bursting bands (indicating population synchronization). To measure the degree of the population synchronization seen in the raster plot, a statistical-mechanical spiking measure $M_s^{(l)}$ of Eq.~(\ref{eq:SM}), based on the ISPSR kernel estimates $R_s^{(l)}(t)$ [$l=1$ (2) corresponds to the stimulated (non-stimulated) case], was introduced by considering the occupation degrees $O_i^{(l)}$ of Eq.~(\ref{eq:OD}) (representing the density of stripes/bands) and the pacing degrees $P_i^{(l)}$ of Eq.~(\ref{eq:PD}) (denoting the smearing of stripes/bands) of the spikes in the stripes/bands \cite{Kim1}: for more details, refer to Appendix \ref{sec:SMSM}. The average occupation degree $\langle \langle O_i^{(l)} \rangle \rangle_r$, the average pacing degree $\langle \langle P_i^{(l)} \rangle \rangle_r$, and the average statistical-mechanical spiking measure $\langle M_s^{(l)} \rangle_r$ ($\langle \cdots \rangle$ and $\langle \cdots \rangle_r$ represent the averages over global cycles and realizations, respectively) are shown  in Figs.~\ref{fig:DREx2}(d1)-\ref{fig:DREx2}(d3), respectively. Moreover, we obtain the cross-correlation function $C_{12}(\tau)$ between $R_s^{(1)}(t)$ and $R_s^{(2)}(t)$ of the two sub-populations:
\begin{equation}
C_{12}(\tau) = \frac{\overline{\Delta R_s^{(1)}(t+\tau) \Delta R_s^{(2)}(t)}} {\sqrt{\overline{ {\Delta R_s^{(1)}}^2(t)}} \sqrt{\overline{ {\Delta R_s^{(2)}}^2(t)}}},
\label{eq:CCF}
\end{equation}
where $\Delta R_s^{(1)}(t) = R_s^{(1)}(t) - \overline{R_s^{(1)}(t)}$, $\Delta R_s^{(2)}(t) = R_s^{(2)}(t) - \overline{R_s^{(2)}(t)}$, and the overline denotes the time average. Then, the cross-correlation measure $M_c$
between the stimulated and the non-stimulated sub-populations is given by the value of $C_{12}(\tau)$ at the zero-time lag:
\begin{equation}
  M_c = C_{12}(0),
\label{eq:CM}
\end{equation}
which corresponds to the Pearson's correlation coefficient for pairs of $[R_s^{(1)}(t), R_s^{(2)}(t)]$ \cite{NR}.
The cross-correlation functions $\langle C_{12}(\tau) \rangle_r$ for various values of $A$ are shown in Figs.~\ref{fig:DREx2}(e1)-\ref{fig:DREx2}(e8), and Figure \ref{fig:DREx2}(f) shows a plot of $\langle M_c \rangle_r$ versus $A$.

We consider the 1st stage [$0 < A < A^*_1(\simeq 83)$] where synchronization enhancement with $\langle D_f \rangle_r>1$ occurs. For small $A$, stimulated RS neurons exhibit spikings which are phase-locked to external AC stimulus $S(t)$ [e.g., see Figs.~\ref{fig:DREx1}(c2) and \ref{fig:DREx2}(a2) for $A=10$]. [In Fig.~\ref{fig:DREx2}, the upper (lower) panels correspond to the stimulated (non-stimulated) case.] Non-stimulated RS neurons also show spikings which are well matched with those of stimulated neurons thanks   to phase-attractive effect of synaptic excitation, as shown in Figs.~\ref{fig:DREx1}(d2) and \ref{fig:DREx2}(a2) for $A=10$. For the case of $A=10$, the widths of stripes in the raster plot of spikes are reduced in comparison with those for $A=0$ [compare Fig.~\ref{fig:DREx1}(a2) with Fig.~\ref{fig:DREx1}(a1)], which implies an increase in the degree of population synchronization. Hence, the oscillating amplitudes of $R_w(t)$, $R_s^{(1)}(t)$, and $R_s^{(2)}(t)$ for $A=10$ become larger than those for $A=0$ [compare Figs.~\ref{fig:DREx1}(b2) and \ref{fig:DREx2}(b2) with Figs.~\ref{fig:DREx1}(b1) and \ref{fig:DREx2}(b1)]. Peaks of $\Delta R_w(t)$, $\Delta R_s^{(1)}(t)$, and $\Delta R_s^{(2)}(t),$ associated with external phase-lockings of both stimulated and non-stimulated RS neurons, appear at the driving frequency
$f_d$ (=7.6 Hz) and its harmonics, as shown in Figs.~\ref{fig:DREx1}(e2) and \ref{fig:DREx2}(c2). In this way, synchronization enhancement occurs, and $\langle D_f \rangle_r$ increases until $A=10$ [see the inset of Fig.~\ref{fig:DREx1}(f)]. However, for $A > 10$ stimulated RS neurons begin to exhibit burstings, in contrast to spiking of non-stimulated RS neurons. Then, due to difference in the type of firings of individual neurons, it is not easy for the spikings of non-stimulated RS neurons to be well matched with burstings of stimulated RS neurons. With increasing $A$, this type of mismatching begins to be gradually intensified, and the degree of population synchronization decreases. Hence, $\langle D_f \rangle_r$ begins to decrease for $A > 10$, as shown in the inset of Fig.~\ref{fig:DREx1}(f).

Eventually, when passing the 1st threshold $A^*_1~(\simeq 83)$, a 2nd stage [$A^*_1 < A < A^*_2(\simeq 1287)$] appears where synchronization suppression with $\langle D_f \rangle_r < 1$ occurs [see Fig.~\ref{fig:DREx1}(f)]. For this case, stimulated RS neurons exhibit burstings which are phase-locked to external AC stimulus $S(t)$. As shown in Figs.~\ref{fig:DREx1}(c3)-\ref{fig:DREx1}(c5) for the membrane potential $v_5(t)$ of the 5th stimulated RS neuron in the stage II, with increasing $A$ the number of spikes in each bursting  increases. On the other hand, non-stimulated RS neurons show persistent spikings which are not well matched with burstings of stimulated neurons [e.g., see Figs.~\ref{fig:DREx1}(d3)-\ref{fig:DREx1}(d5) for the membrane potential $v_{20}(t)$ of the 20th non-stimulated RS neuron]. As an example, consider the case of $A=150$. Stimulated RS neurons exhibit burstings, each of which consists of two spikes, as shown in Fig.~\ref{fig:DREx1}(c3). These burstings are synchronized, and hence a pair of vertical trains (composed of synchronized spikes in burstings) appear successively in the raster plot of spikes, as shown in the upper panel of Fig.~\ref{fig:DREx2}(a3). On the other hand, spiking stripes of non-stimulated RS neurons are smeared in a zigzag way between the synchronized vertical bursting trains (i.e., the pacing degree between spikes of non-stimulated RS neruons is reduced) [see the lower panel of Fig.~\ref{fig:DREx2}(a3)]. This zigzag pattern (indicating local clustering of spikes) in the smeared stripes seems to appear because the Watts-Strogatz SWN with $p=0.2$ has a relatively high clustering coefficient (denoting cliquishness of a typical neighborhood in the network) \cite{SWN-Kim}. These zigzag smeared stripes also appear in a nearly regular way with the driving frequency $f_d$, like the case of vertical bursting trains of stimulated RS neurons [see Fig.~\ref{fig:DREx2}(a3)]. Hence, both cases of stimulated and non-stimulated RS neurons are phase-locked to external AC stimulus, although they are mismatched (i.e., phase-shifted). Peaks of $\Delta R_w(t)$, $\Delta R_s^{(1)}(t)$, and $\Delta R_s^{(2)}(t)$, corresponding to these external phase-lockings, appear at the driving frequency $f_d$ and its harmonics, as shown in Figs.~\ref{fig:DREx1}(e3) and \ref{fig:DREx2}(c3). Phase-shifted mixing of synchronized vertical bursting trains (of stimulated RS neurons) and zigzag smeared spiking stripes (of non-stimulated RS neurons) leads to decrease in the degree of population synchronization. Consequently, the amplitudes of $R_w(t)$ and $R_s^{(2)}(t)$ for $A=150$ are smaller than those for $A=10$ [compare Fig.~\ref{fig:DREx1}(b3) and Fig.~\ref{fig:DREx2}(b3) with Fig.~\ref{fig:DREx1}(b2) and Fig.~\ref{fig:DREx2}(b2)], and synchronization suppression (with $\langle D_f \rangle_r<1$) occurs [see Fig.~\ref{fig:DREx1}(f)]. As $A$ is further increased, more number of synchronized vertical busting trains (phase-locked to external stimulus) appear successively in the raster plot of spikes, because each bursting of stimulated RS neurons consists of more number of spikes. Zigzag smearing of spiking stripes of non-stimulated RS neurons becomes intensified (i.e., the pacing degree between spikes of non-stimulated RS neurons becomes worse), although they are phase-locked to external AC stimulus. These bursting trains and smeared spiking stripes are still phase-shifted. In this way, with increasing $A$, the degree of population synchronization is decreased mainly due to smearing of spiking stripes, and eventually a minimum ($\simeq 0.7937$) of $\langle D_f \rangle_r$ occurs for $A = A_{min}^{(1)} (\simeq 398)$, as shown in Fig.~\ref{fig:DREx1}(f). An example near this minimum is given for the case of $A=400$. A quadruple of vertical trains (consisting of synchronized spikes in burstings of stimulated RS neurons) and zigzag smeared spiking stripes of non-stimulated RS neurons appear successively in the raster plot of spikes, as shown in Fig.~\ref{fig:DREx2}(a4). Both of them are phase-locked to external stimulus, but they are more phase-shifted. Peaks of $\Delta R_w(t)$, $\Delta R_s^{(1)}(t)$, and $\Delta R_s^{(2)}(t)$, related to external phase-lockings for both cases of stimulated and non-stimulated RS neurons, also appear at the driving frequency $f_d$ and its harmonics [see Figs.~\ref{fig:DREx1}(e4) and \ref{fig:DREx2}(c4)]. Furthermore, the spiking stripes of non-stimulated RS neurons are much more smeared in a zigzag way when compared with the case of $A=150$ [compare Fig.~\ref{fig:DREx2}(a4) with Fig.~\ref{fig:DREx2}(a3)]. Hence, the amplitudes of $R_w(t)$ and $R_s^{(2)}(t)$ become smaller than those for $A=150$ (i.e., the degree of population synchronization is more reduced) [compare Figs.~\ref{fig:DREx1}(b4) and \ref{fig:DREx2}(b4) with Figs.~\ref{fig:DREx1}(b3) and \ref{fig:DREx2}(b3)].

However, with further increase in $A$ from $A_{min}^{(1)}$, synchronized burstings of stimulated RS neurons are more developed. Moreover, widths of zigzag smeared spiking stripes of non-stimulated RS neurons become gradually reduced [i.e., the degree of mismatching (phase-shift) between the stimulated and the non-stimulated sub-populations becomes decreased]. A constructive effect of $S(t)$ (resulting from a phase-attractive synaptic excitation) seems to appear effectively. Consequently, the degree of population synchronization begins to increase (i.e., $\langle D_f \rangle_r$ starts to grow). As an example, we consider the case of $A=800$. Both the bursting bands (composed of spikes in burstings of the stimulated RS neurons) and the spiking stripes of non-stimulated RS neurons, phase-locked to external AC stimulus, appear successively in the raster plots of spikes, as shown in Fig.~\ref{fig:DREx2}(a5). When compared with the case of $A=400$, the bursting bands are more developed, and the degree of zigzag smearing of spiking stripes is reduced [compare Fig.~\ref{fig:DREx2}(a5) with Fig.~\ref{fig:DREx2}(a4)]. Both the bursting bands and the smeared stripes are phase-locked to external AC stimulus, and their phase-shift is reduced. Peaks of $\Delta R_w(t)$, $\Delta R_s^{(1)}(t)$, and $\Delta R_s^{(2)}(t)$, related to external phase-lockings for both cases of stimulated and non-stimulated RS neurons, also appear at the driving frequency $f_d$ and its harmonics [see Figs.~\ref{fig:DREx1}(e5) and \ref{fig:DREx2}(c5)]. Hence, the amplitudes of $R_s^{(1)}(t)$ and $R_s^{(2)}(t)$ become larger than those for $A=400$ [compare Fig.~\ref{fig:DREx2}(b5) with Fig.~\ref{fig:DREx2}(b4)], which results in the increase in the amplitude of $R_w(t)$ [compare Fig.~\ref{fig:DREx1}(b5) with Fig.~\ref{fig:DREx1}(b4)]. As a result, the degree of population synchronization is larger than that for $A=400$. In this way, with increasing $A$ from $A_{min}^{(1)}$ the dynamical factor $\langle D_f \rangle_r$ is increased. Eventually, when passing the 2nd threshold $A^*_2~(\simeq 1287)$, $\langle D_f \rangle_r$ passes the unity, and a 3rd stage appears, where synchronization enhancement with $\langle D_f \rangle_r>1$ reappears thanks to a phase-attractive effect of synaptic excitation [see Fig.~\ref{fig:DREx1}(f)]. As examples, we consider the cases of $A=2000$, 5000, and $10^4$. As $A$ is increased in this 3rd stage, burstings of RS neurons are more developed [e.g., see Figs.~\ref{fig:DREx1}(c6)-\ref{fig:DREx1}(c8)], and non-stimulated RS neurons also begin to fire burstings for sufficiently large $A$ [e.g., see Figs.~\ref{fig:DREx1}(d7)-\ref{fig:DREx1}(d8)]. Then, synchronized bursting bands of stimulated RS neurons are more and more intensified, as shown in Figs.~\ref{fig:DREx2}(a6)-\ref{fig:DREx2}(a8). Moreover, ``firing'' bands, composed of spikings/burstings of non-stimulated RS neurons, become matched well with bursting bands of stimulated RS neurons [see Figs.~\ref{fig:DREx2}(a6)-\ref{fig:DREx2}(a8)]: the matching degree also increases with $A$. Peaks of $\Delta R_w(t)$, $\Delta R_s^{(1)}(t)$, and $\Delta R_s^{(2)}(t)$, related to external phase-lockings for both cases of stimulated and non-stimulated RS neurons, appear at the driving frequency $f_d$ and its harmonics [see Figs.~\ref{fig:DREx1}(e6)-\ref{fig:DREx1}(e8) and Figs.~\ref{fig:DREx2}(c6)-\ref{fig:DREx2}(c8)]. Consequently, with increasing $A$ the amplitudes of both $R_s^{(1)}(t)$ and $R_s^{(2)}(t)$ are increased, as shown in Figs.~\ref{fig:DREx2}(b6)-\ref{fig:DREx2}(b8), which also leads to increase in $R_w(t)$ [see Figs.~\ref{fig:DREx1}(b6)-\ref{fig:DREx1}(b8)]. In this way, $\langle D_f \rangle_r$ increases monotonically with $A$ and synchronization enhancement occurs in the 3rd stage, as shown in Fig.~\ref{fig:DREx1}(f).

By varying $A$, we also characterize population synchronization in each of the stimulated and the non-stimulated sub-populations in terms of the average occupation degree $\langle \langle O_i^{(l)} \rangle \rangle_r$, the average pacing degree $\langle \langle P_i^{(l)} \rangle \rangle_r$, and the statistical-mechanical spiking measure $\langle M_s^{(l)} \rangle_r$; $l=1$ and 2 correspond to the stimulated and the non-stimulated cases, respectively. Plots of $\langle \langle O_i^{(l)} \rangle \rangle_r$, $\langle \langle P_i^{(l)} \rangle \rangle_r$, and $\langle M_s^{(l)} \rangle_r$ versus $A$ are shown in Figs.~\ref{fig:DREx2}(d1)-\ref{fig:DREx2}(d3), respectively. As $A$ is increased, external phase lockings of spikings or burstings of stimulated RS neurons are more and more enhanced, as shown in Figs.~\ref{fig:DREx2}(a1)-\ref{fig:DREx2}(a8). Hence, the stimulated RS neurons exhibit full synchronization with $\langle \langle O_i^{(1)} \rangle \rangle_r =1,$ independently of $A$ because every stimulated RS neuron makes a firing in each spiking stripe or bursting band [corresponding to each global cycle of $R_s^{(1)}(t)$]. These fully synchronized spikes also show high average pacing degree $\langle \langle P_i^{(1)} \rangle \rangle_r$. For $0 < A < 10,$ $\langle \langle P_i^{(1)} \rangle \rangle_r$ increases monotonically from 0.967 to 0.998 because smearing of spiking stripes (i.e. width of spiking stripes) becomes reduced [see the left inset of Fig.~\ref{fig:DREx2}(d2)]. For $A>10$ bursting bands appear, at first their widths increase, but eventually they become saturated for large $A$ [see Figs.~\ref{fig:DREx2}(a3)-\ref{fig:DREx2}(a8)]. Hence, for $A>10$ $\langle \langle P_i^{(1)} \rangle \rangle_r$ begins to decrease, but it seems to approach a limit value $(\simeq 0.82$). Consequently, the average spiking measure $\langle M_s^{(1)} \rangle_r$ (given by taking into consideration both the occupation and the pacing degrees) exhibit the same behaviors with $A$ as $\langle \langle P_i^{(1)} \rangle \rangle_r$ because $\langle \langle O_i^{(1)} \rangle \rangle_r =1.$ We next consider the non-stimulated case. Non-stimulated RS neurons also exhibit full synchronization with $\langle \langle O_i^{(2)} \rangle \rangle_r =1,$ independently of $A$. However, the average pacing degree $\langle \langle P_i^{(2)} \rangle \rangle_r$ varies with $A$, differently from the stimulated case. For $0 < A <10$, spikings of non-stimulated RS neurons are well matched with those of stimulated RS neurons thanks to phase-attractive effect of synaptic excitation, and hence the average pacing degree $\langle \langle P_i^{(2)} \rangle \rangle_r $ increases monotonically from $0.973$ to $0.989$ [see the right inset of Fig.~\ref{fig:DREx2}(d2)]. However, for $A>10$ it is not easy for spikings of non-stimulated neurons to be well matched with burstings of stimulated neurons because of different firing type. Hence, zigzag smearing occurs in the spiking stripes of non-stimulated neurons, and it is enhanced with $A$. Due to such developed zigzag smearing, $\langle \langle P_i^{(2)} \rangle \rangle_r$ decreases with $A$, and it arrives at its minimum ($\simeq 0.483$) for $A \simeq 391$. As $A$ is further increased from the minimum point, zigzag smearing begins to be gradually reduced thanks to a constructive effect of $S(t)$ (coming from the phase-attractive synaptic excitation). As a result, $\langle \langle P_i^{(2)} \rangle \rangle_r$ starts to increase, and its value becomes large for large $A$ (e.g.,  $\langle \langle P_i^{(2)} \rangle \rangle_r \simeq 0.65$ for $A=10^4$). The average spiking measure $\langle M_s^{(2)} \rangle_r$ also show the same behaviors with $A$ as $\langle \langle P_i^{(2)} \rangle \rangle_r$ because $\langle \langle O_i^{(2)} \rangle \rangle_r =1.$

Finally, to examine the matching degree between the stimulated and the non-stimulated sub-populations, we obtain the cross-correlation functions $\langle C_{12}(\tau) \rangle_r$ between $R_s^{(1)}(t)$ and $R_s^{(2)}(t)$ of the two sub-populations, which are shown for various values of $A$ in Figs.~\ref{fig:DREx2}(e1)-\ref{fig:DREx2}(e8). A plot of the cross-correlation measure $\langle M_c \rangle_r$ [given by $C_{12}(0)$] versus $A$ is also shown in Fig.~\ref{fig:DREx2}(f). Perfect cross-correlation with $\langle M_c \rangle_r = 1$ occurs in the range of $0<A<10$ where $\langle D_f \rangle_r$ increases monotonically from 1 to its maximum ($\simeq 1.095$) at $A=10$ [see the inset in Fig.~\ref{fig:DREx1}(f)]. In the remaining region ($10<A<A^*_1$) of the 1st stage, $\langle M_c \rangle_r$ decreases slowly, but it still indicates strong cross-correlation with $\langle M_c \rangle_r > 0.97$. This type of perfect/strong cross-correlation induces phase-attractive effect between the stimulated and the non-stimulated sub-populations, and hence synchronization enhancement occurs in the stage I. However, in the first part of the 2nd stage $\langle M_c \rangle_r$ decreases very rapidly to its minimum for $A \simeq 403$ (which is nearly the same as $A_{min}^{(1)} (\simeq 398)$ for the minimum of $\langle D_f \rangle_r$), mainly because of the different firing type of the stimulated RS neurons (bursting) and the non-stimulated FS interneurons (spiking). Due to sudden decrease in the cross-correlation, $\langle D_f \rangle_r$ also decreases from 1, and synchronization suppression occurs. After passing the minimum point ($A \simeq 403$), $\langle M_c \rangle_r$ begins to increase gradually with $A$, thanks to a phase-attractive effect of the excitatory coupling. Consequently, in the latter part of the 2nd stage (with $\langle D_f \rangle_r <1)$ $\langle D_f \rangle_r$ increases monotonically with $A$, and eventually when passing the 2nd threshold $A^*_2(\simeq 1287)$ $\langle D_f \rangle_r$ passes the unity. Thus, the 3rd stage appears, and synchronization enhancement reoccurs.

\subsection {Dynamical Responses of Inhibitory Synchronization to An External Time-Periodic Stimulus}
\label{subsec:DRIn}
\begin{figure}
\includegraphics[width=0.6\columnwidth]{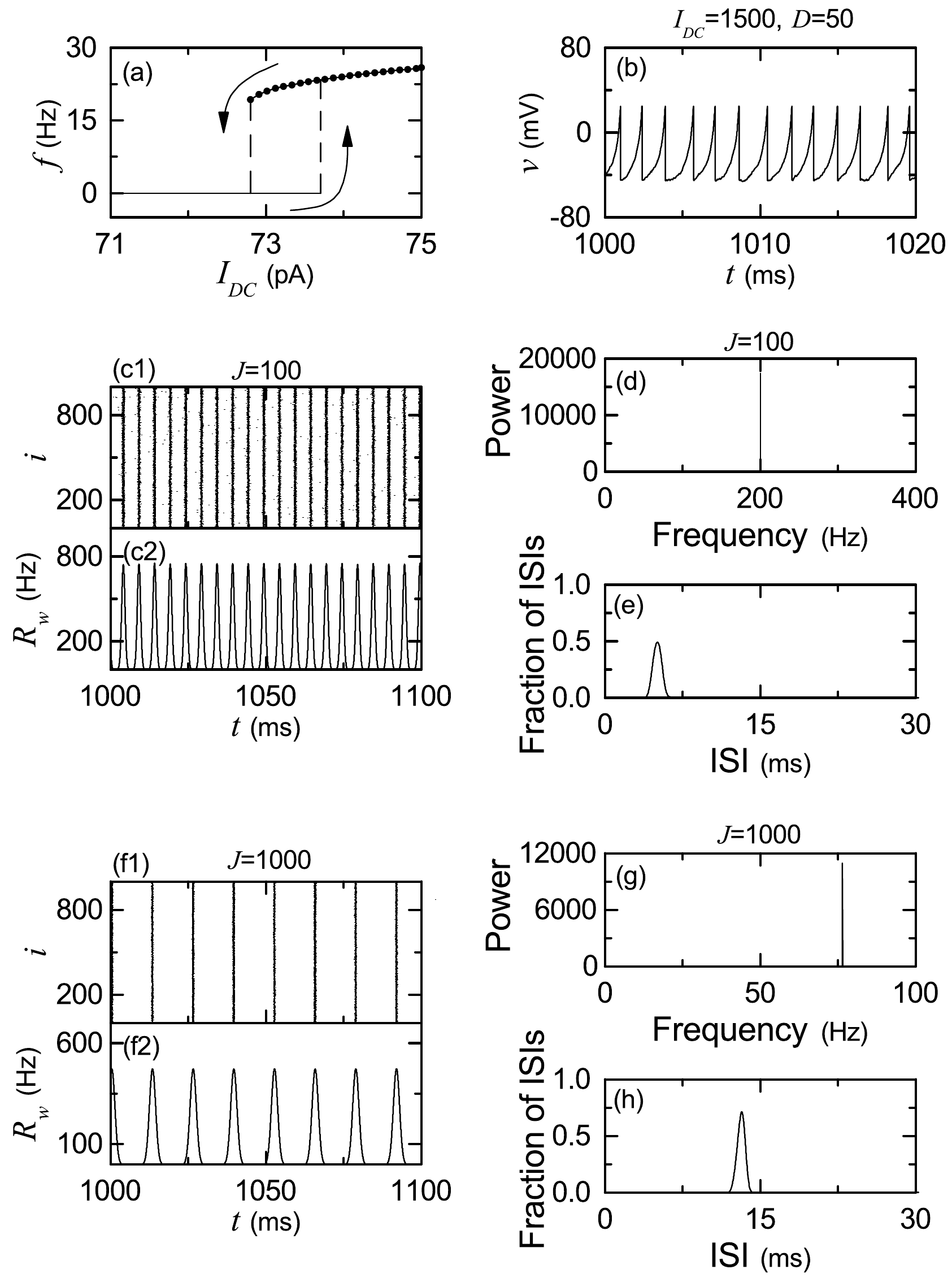}
\linespread{1.1}
\caption{Inhibitory synchronization in the Watts-Strogatz SWN of $N (=10^3)$ Izhikevich FS interneurons for $M_{syn}=50$ and $p=0.2$. Single Izhikevich FS interneuron:(a) plot of the mean firing rate $f$ vs. the external DC current $I_{DC}$ for $D=0$ and (b) time series of the membrane potential $v$ for $I_{DC}=1500$ and $D=50$. Coupled Izhikevich FS interneurons for $I_{DC}=1500$, $D=50$, and $J=100$: (c1) raster plot of spikes, (c2) plot of the instantaneous whole-population spike rate (IWPSR) kernel estimate $R_w(t)$ versus $t$, (d) one-sided power spectrum of $\Delta R_w(t) [=R_w(t) - \overline{R_w(t)}]$ (the overbar represents the time average) with mean-squared amplitude normalization, and (e) inter-spike interval (ISI) histogram. Coupled Izhikevich FS interneurons for $I_{DC}=1500$, $D=50$, and $J=1000$: (f1) raster plot of spikes, (f2) plot of the IWPSR kernel estimate $R_w(t)$ versus $t$, (g) one-sided power spectrum of $\Delta R_w(t) [=R_w(t) - \overline{R_w(t)}]$ with mean-squared amplitude normalization, and (h) ISI histogram. The band widths of the Gaussian kernel estimates for the IWPSR $R_w(t)$ are 0.5 ms and 1.0 ms for $J=100$ and 1000, respectively. Each power spectrum is obtained via 30 realizations [$2^{16} (=65536)$ data points are used in each realization], each ISI histogram is also obtained through 30 realizations ($5 \times 10^7$ ISIs are used in each realization), and the bin size for the histogram is 0.5 ms.
}
\label{fig:FS}
\end{figure}

We consider an inhibitory Watts-Strogatz SWN composed of $N (=10^3)$ Izhikevich FS interneurons. Figure \ref{fig:FS}(a) shows a plot of the firing frequency $f$ versus the external DC current $I_{DC}$ for a single Izhikevich FS interneuron in the absence of noise ($D=0$). The Izhikevich FS interneuron exhibits a jump from a resting state to a spiking state via subcritical Hopf bifurcation at a higher threshold $I_{DC,h} (\simeq 73.7)$ by absorbing an unstable limit cycle born through a fold limit cycle bifurcation for a lower threshold $I_{DC,l} (\simeq 72.8)$. Hence, the Izhikevich FS interneuron exhibits type-II excitability because it begins to fire with a non-zero frequency \cite{Izhi3,Ex1,Ex2}. As $I_{DC}$ is increased from $I_{DC,h}$, the firing frequency $f$ increases monotonically. Here, we consider a suprathreshold case of $I_{DC}=1500$ in the presence of noise with $D=50$ for which a time series of the membrane potential $v$ with an oscillating frequency $f \simeq 635$ Hz is shown in Fig.~\ref{fig:FS}(b). We consider two coupling cases of $J=100$ and 1000 to study the effect of coupling strength $J$ on the dynamical responses. Full synchronization for $J=100$ is well shown in the raster plot of spikes in Fig.~\ref{fig:FS}(c1). For this case, the IWPSR kernel estimate $R_w(t)$ exhibits a regular oscillation with a fast population frequency $f_p~(\simeq 200$ Hz) [see the peak in the power spectrum of $\Delta R_w(t)$ in Fig.~\ref{fig:FS}(d)]. The ISI histogram for individual interneurons is also shown in Fig.~\ref{fig:FS}(e). The ensemble-averaged ISI $\langle ISI \rangle$ is 5.0 ms, and hence the ensemble-averaged MFR $\langle f_i \rangle$ of individual interneurons (corresponding to the reciprocal of $\langle ISI \rangle$) is 200 Hz, which is the same as $f_p$. For a strong-coupling case of $J=1000$, the raster plot of spikes and the IWPSR kernel estimate $R_w(t)$ in Figs.~\ref{fig:FS}(f1) and \ref{fig:FS}(f2) show full synchronization well. The population frequency $f_p$ of $R_w(t)$ is 76 Hz [see the peak in the power spectrum of $\Delta R_w(t)$ in Fig.~\ref{fig:FS}(g)], which is smaller than that for $J=100$ because of strong inhibition. The ensemble-averaged ISI $\langle ISI \rangle$ in Fig.~\ref{fig:FS}(h) is 13.1 ms which is longer than that for $J=100$. Hence, the ensemble-averaged MFR $\langle f_i \rangle$ of individual interneurons is 76 Hz, which is also the same as $f_p$.

\subsubsection{Small-Coupling Case of $J=100$}
\label{subsubsec:SJ}
\begin{figure}
\includegraphics[width=0.8\columnwidth]{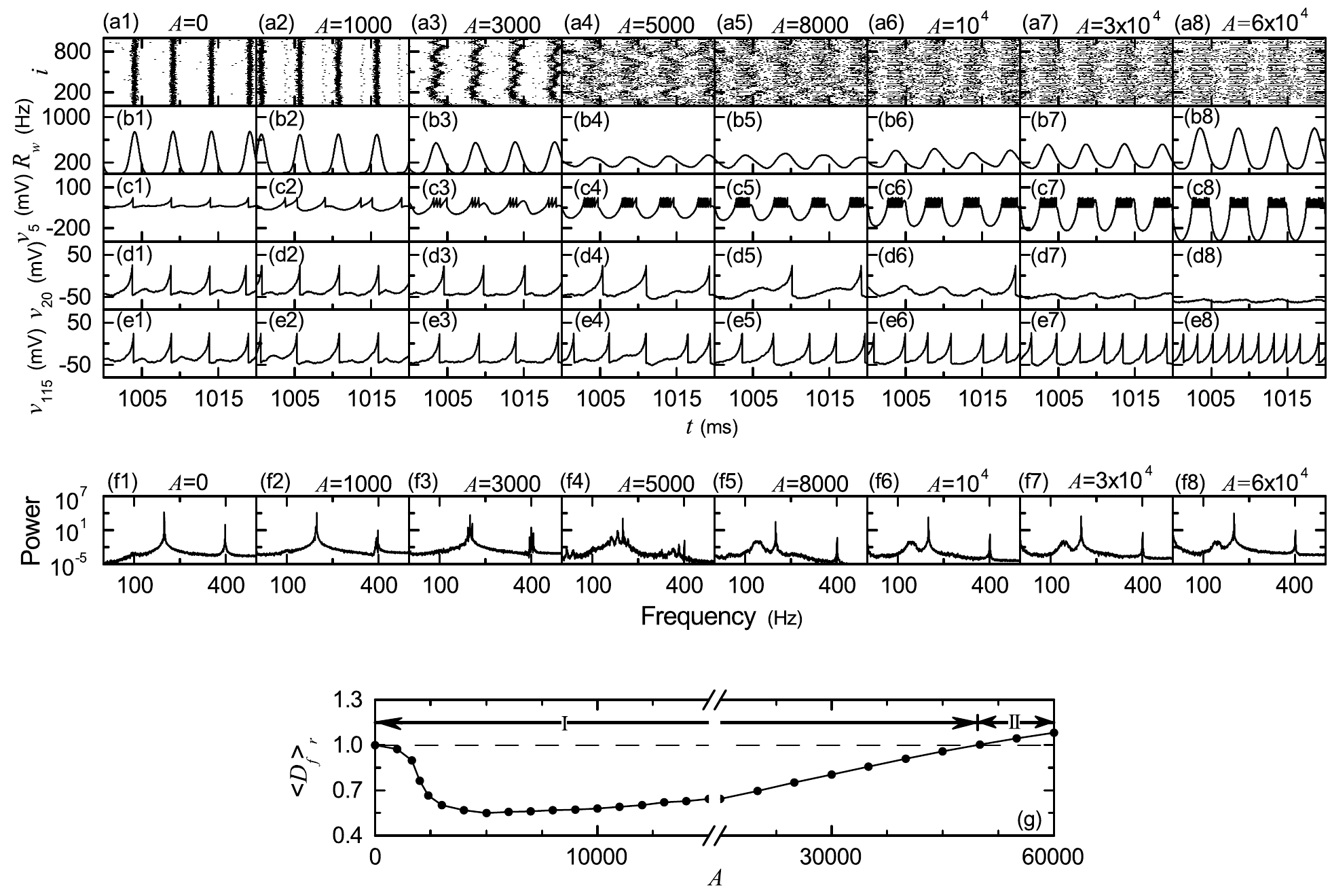}
\caption{Dynamical response when an external time-periodic stimulus $S(t)$ is applied to 50 randomly-chosen Izhikevich FS interneurons in the case of inhibitory synchronization for $J=100$. We vary the driving amplitude $A$ for a fixed driving angular frequency $\omega_d$ (=1.26 rad/ms). Raster plots of spikes, instantaneous whole-population spike rate (IWPSR) kernel estimates $R_w(t)$, membrane potentials $v_5(t)$ of the stimulated 5th FS interneuron,  membrane potentials $v_{20}(t)$ of the major non-stimulated 20th FS interneuron, and membrane potentials $v_{115}(t)$ of the minor  non-stimulated 115th FS interneuron are shown for various values of $A$ in (a1)-(a8), (b1)-(b8), (c1)-(c8), (d1)-(d8), and (e1)-(e8), respectively. One-sided power spectra of $\Delta R_w(t) [=R_w(t) - \overline{R_w(t)}]$ (the overbar represents the time average) with mean-squared amplitude normalization are also given in (f1)-(f8). The band width of the Gaussian kernel estimate for each IWPSR $R_w(t)$ is 0.5 ms, and each power spectrum is obtained via 30 realizations [$2^{16} (=65536)$ data points are used in each realization].
(g) Plot of dynamical response factor $\langle D_f \rangle_r$ versus $A$, where I and II represent the 1st (synchronization suppression) and the 2nd (synchronization enhancement) stages, respectively. Here, $\langle \cdots \rangle_r$ represents an average over 30 realizations. Averaging time for $D_f$ in each realization is $3 \times 10^4$ ms.
}
\label{fig:DRIh1}
\end{figure}

We first consider the case of $J=100$. We apply an external time-periodic AC stimulus $S(t)$ to $N_s(=50)$ randomly-selected Izhikevich FS interneurons by fixing the driving angular frequency as $\omega_d (=2 \pi f_d)$ =1.26 rad/ms ($f_d = \langle f_i \rangle$ =200 Hz), and investigate the dynamical response of inhibitory full synchronization by varying the driving amplitude $A$. Figures \ref{fig:DRIh1}(a1)-\ref{fig:DRIh1}(a8) show raster plots of spikes for various values of $A$. Population synchronization may be well seen in these raster plots of spikes. The IWPSR kernel estimates $R_w(t),$ exhibiting oscillatory behaviors, are shown in Figs.~\ref{fig:DRIh1}(b1)-\ref{fig:DRIh1}(b8), and the power spectra of $\Delta R_w(t)$ are also given in Figs.~\ref{fig:DRIh1}(f1)-\ref{fig:DRIh1}(f8). In addition, times series of membrane potentials of individual FS interneurons are given for various values of $A$. The time series of $v_5(t)$ of the 5th stimulated FS interneuron are shown in Figs.~\ref{fig:DRIh1}(c1)-\ref{fig:DRIh1}(c8). For the non-stimulated case, there are two types of FS interneurons, depending on their synaptic connections. Many non-stimulated FS interneurons (i.e., major non-stimulated FS interneurons) which have synaptic connections with fast-firing stimulated FS interneurons fire slowly due to increased inhibition. On the other hand, a small number of non-stimulated FS interneurons (i.e., minor non-stimulated FS interneurons) which have no direct synaptic connections with stimulated FS interneurons receive synaptic inputs from major slowly-firing non-stimulated FS interneurons, and hence MFRs of minor non-stimulated FS interneurons become fast due to decreased inhibition. Figures \ref{fig:DRIh1}(d1)-\ref{fig:DRIh1}(d8) show the time series of $v_{20}(t)$ of the 20th major slowly-firing non-stimulated FS interneuron, while Figs.~\ref{fig:DRIh1}(e1)-\ref{fig:DRIh1}(e8) show the time series of $v_{115}(t)$ of the 115th minor fast-firing non-stimulated FS interneuron. A plot of the dynamical factor $\langle D_f \rangle_r$ versus $A$ is given in Fig.~\ref{fig:DRIh1}(g). Two stages are thus found to appear. Synchronization suppression ($\langle D_f \rangle_r <1$) and synchronization enhancement
($\langle D_f \rangle_r >1$) occur in the 1st (I) stage ($0< A < A^*_3$) and the 2nd (II) stage ($A > A^*_3$), respectively, where $A^*_3 \simeq 49699$. Examples are given for various values of $A$; 1st stage ($A=1000$, 3000, 5000, 8000, $10^4$, and $3 \times 10^4$) and 2nd stage ($A=6 \times 10^4$).

\begin{figure}
\includegraphics[width=0.8\columnwidth]{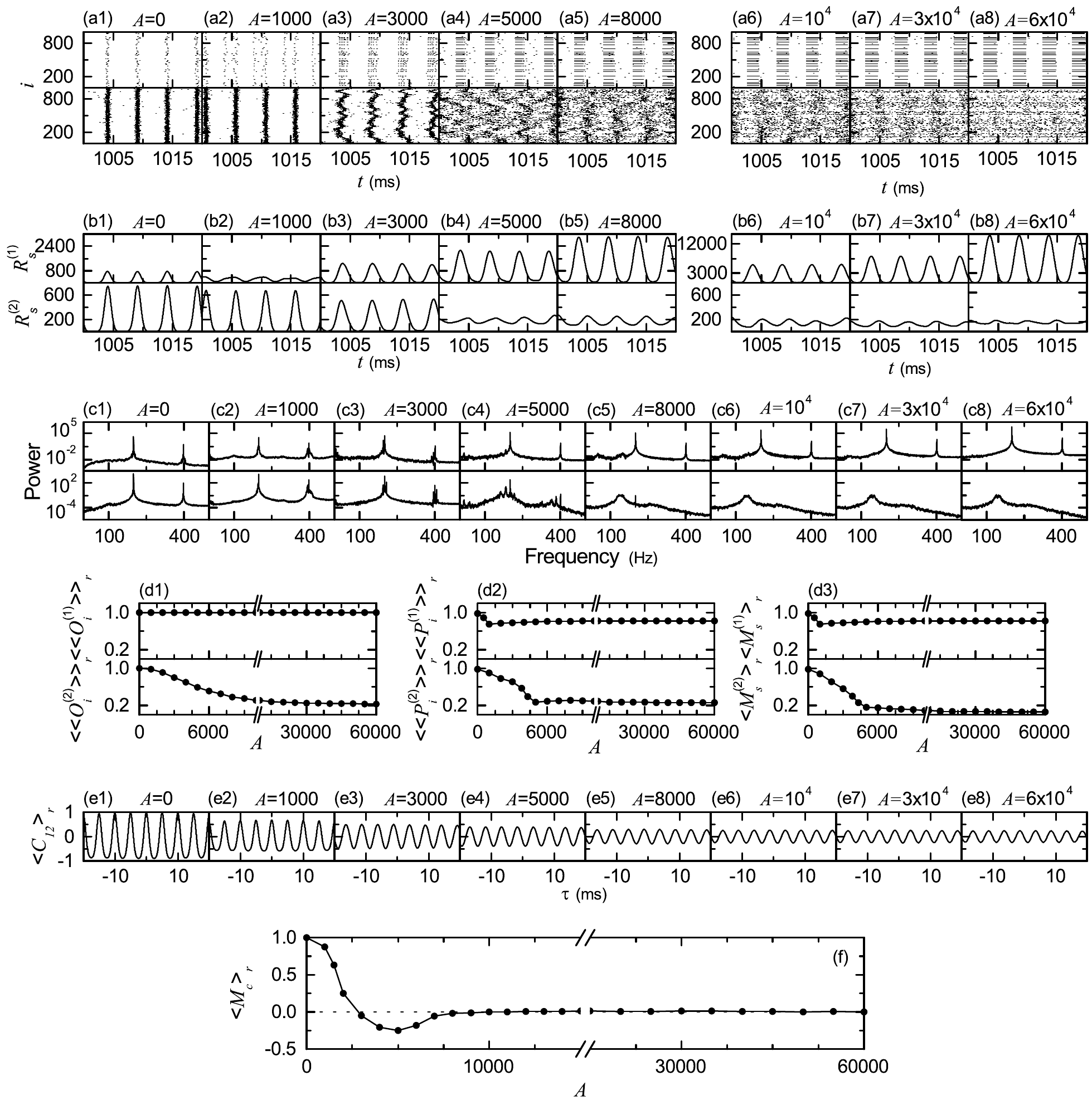}
\linespread{0.88}
\caption{Dynamical responses for the case of Fig.~\ref{fig:DRIh1} in both the stimulated and the non-stimulated sub-populations and a cross-correlation measure $\langle M_c \rangle_r$ between the dynamics of the two sub-populations. Raster plot of spikes, instantaneous sub-population spike rates (ISPSRs) $R_s^{(1)}$ and $R_s^{(2)}$ [the superscript 1 (2) corresponds to the stimulated (non-stimulated) case], and one-sided power spectra of $\Delta R_s^{(1)}(t) [=R_s^{(1)}(t) - \overline{R_s^{(1)}(t)}]$ and  $\Delta R_s^{(2)}(t) [=R_s^{(2)}(t) - \overline{R_s^{(2)}(t)}]$ (the overbar represents the time average) with mean-squared amplitude normalization in the stimulated and the non-stimulated sub-populations are shown for various values of $A$ in (a1)-(a8), (b1)-(b8), and (c1)-(c8), respectively: the upper (lower) panels denote those for the stimulated (non-stimulated) case. The band width of the Gaussian kernel estimate for each ISPSR is 0.5 ms, and each power spectrum is obtained via 30 realizations [$2^{16} (=65536)$ data points are used in each realization]. Plots of the average occupation degree $\langle \langle O_i^{(l)} \rangle \rangle_r$, the average pacing degree $\langle \langle P_i^{(l)} \rangle \rangle_r$, and the statistical-mechanical spiking measure $\langle M_s^{(l)} \rangle_r$ versus $A$ are shown in (d1)-(d3), respectively; $l=1$ (2) corresponds to the stimulated (non-stimulated) case. Here, $\langle O_i^{(l)} \rangle$, $\langle P_i^{(l)} \rangle$, and $M_s^{(l)}$ are obtained by following the $3 \times 10^3$ stripes/bands in the raster plot of spikes in each realization, and $\langle \cdots \rangle_r$ denotes an average over 30 realizations. Plots of cross-correlation functions $\langle C_{12}(\tau) \rangle_r$ versus $\tau$ are also given in (e1)-(e8). The number of data used for the calculation of each temporal cross-correlation function $C_{12}(\tau)$ is $2^{16} (=65536)$ in each realization, and $\langle C_{12}(\tau) \rangle_r$ is obtained through an average over 30 realizations. (f) Plot of the cross-correlation measure $\langle M_c \rangle_r$ versus $A$; $\langle M_c \rangle_r$ is obtained via an average over 30 realizations.
}
\label{fig:DRIh2}
\end{figure}

As in the above excitatory case, we make more detailed analysis of dynamical responses by decomposing the whole population of FS interneurons into two sub-populations of the stimulated and the non-stimulated FS interneurons. Dynamical responses in these two sub-populations are shown well in Fig.~\ref{fig:DRIh2}. Raster plots of spikes, ISPSR kernel estimates $R_s^{(1)}(t)$ and $R_s^{(2)}(t)$ [the superscript 1 (2) corresponds to the stimulated (non-stimulated) case], and power spectra of $\Delta R_s^{(1)}(t)$ and $\Delta R_s^{(2)}(t)$ in the stimulated and the non-stimulated sub-populations are shown in Figs.~\ref{fig:DRIh2}(a1)-\ref{fig:DRIh2}(a8), Figs.~\ref{fig:DRIh2}(b1)-\ref{fig:DRIh2}(b8), and Figs.~\ref{fig:DRIh2}(c1)-\ref{fig:DRIh2}(c8), respectively: the upper (lower) panels in these figures represent those for the stimulated (non-stimulated) case. For characterization of population synchronization in each of the stimulated and the non-stimulated sub-populations, the average occupation degree $\langle \langle O_i^{(l)} \rangle \rangle_r$, the average pacing degree $\langle \langle P_i^{(l)} \rangle \rangle_r$, and the average statistical-mechanical spiking measure $\langle M_s^{(l)} \rangle_r$ are given  in Figs.~\ref{fig:DRIh2}(d1)-\ref{fig:DRIh2}(d3), respectively; $l=1$ (2) represents the stimulated (non-stimulated) case. The cross-correlation functions $\langle C_{12}(\tau) \rangle_r$ between $R_s^{(1)}(t)$ and $R_s^{(2)}(t)$ of the two sub-populations are also shown for various values of $A$ in Figs.~\ref{fig:DRIh2}(e1)-\ref{fig:DRIh2}(e8). Figure \ref{fig:DRIh2}(f) shows a plot of the cross-correlation measure $\langle M_c \rangle_r$ [given by $C_{12}(0)$] versus $A$.

As $A$ is increased from 0 and passes a threshold, stimulated FS interneurons begin to exhibit burstings, as shown in Figs.~\ref{fig:DRIh1}(c2)-\ref{fig:DRIh1}(c8), and the number of spikings in each bursting increases with $A$. These burstings are phase-locked to external stimulus $S(t)$, which are intensified with increasing $A$ [see Figs.~\ref{fig:DRIh2}(a2)-\ref{fig:DRIh2}(a8)]. Consequently, as $A$ is increased, the amplitude of $R_s^{(1)}(t)$ also increases, as shown in Figs.~\ref{fig:DRIh2}(b2)-\ref{fig:DRIh2}(b8). Peaks in the power spectrum of $\Delta R_s^{(1)}(t)$, associated with the external phase lockings, appear at the driving frequency $f_d $ (=200 Hz) and its harmonics [see the upper panels of Figs.~\ref{fig:DRIh2}(c2)-\ref{fig:DRIh2}(c8)]. This kind of external phase lockings of stimulated FS interneurons are similar to those for the case of excitatory coupling. However, the external stimulus $S(t)$ makes a destructive effect on the sub-population of non-stimulated FS interneurons, in contrast to the excitatory case (where a constructive effect of $S(t)$, resulting from the phase-attractive synaptic excitation, leads to external phase lockings of non-stimulated RS neurons).

In the presence of burstings of stimulated FS interneurons, spikings of non-stimulated FS interneurons cannot be well matched with burstings of stimulated FS interneurons, because of difference in the type of firings of individual neurons [e.g., see Fig.~\ref{fig:DRIh2}(a2) for $A=1000$]. However, these spiking stripes of non-stimulated FS interneurons are also phase-locked to external stimulus, although they are phase-shifted from the vertical bursting trains of the stimulated FS interneurons. Peaks in the power spectrum of $\Delta R_s^{(2)}(t)$, related to the external phase lockings, appear at the driving frequency $f_d $ (=200 Hz) and its harmonics [see the lower panel of Figs.~\ref{fig:DRIh2}(c2)]. As $A$ is further increased, a destructive effect of $S(t)$, resulting from repulsive synaptic inhibition, becomes intensified. Hence, zigzag smearing pattern appears in their spiking stripes, as shown in Fig.~\ref{fig:DRIh2}(a3) for $A=3000$. As explained in the excitatory case, such zigzag pattern in the smeared stripes seems to appear because the Watts-Strogatz SWN with $p=0.2$ has a relatively high clustering coefficient \cite{SWN-Kim}. Furthermore, major non-stimulated FS interneurons begin to exhibit intermittent and stochastic spikings (i.e., stochastic spike skipping) \cite{GR,Longtin1,Longtin2}. Due to the stochastic spike skipping, the original full synchronization (where all the non-stimulated FS interneurons fire spikings in each spiking stripe) in the non-stimulated sub-population begins to break up, and a sparse synchronization (where only some fraction of non-stimulated FS interneurons fire spikings in each spiking stripe) starts to appear (i.e., sparse spiking stripes begin to appear) \cite{Sparse,SWN-Kim}. [However, the degree of sparseness for $A=3000$ is relatively low, and hence no skippings are found in $v_{20}(t)$ of the 20th major non-stimulated FS interneuron for a short time interval of 20 ms in Fig.~\ref{fig:DRIh1}(d3).] In this way, with increasing $A$ the mismatching degree between the stimulated and the non-stimulated sub-populations is increased, although both the bursting bands of stimulated FS interneurons and the zigzag smeared sparse spiking stripes of non-stimulated FS interneurons are phase locked to external stimulus. Due to increased zigzag smearing, peaks at the driving frequency $f_d$ and its harmonics for $A=3000$ become more broad than those for $A=1000$ [compare Fig.~\ref{fig:DRIh2}(c3) with Fig.~\ref{fig:DRIh2}(c2)]. The effect of zigzag smearing and stochastic spike skipping in the non-stimulated sub-population is more dominant when compared with the enhanced external phase lockings in the stimulated sub-population. Hence, the overall degree of population synchronization in the whole population becomes worse. As a result, for the case of $A=3000$, the amplitudes of $R_s^{(2)}(t)$ and $R_w(t)$ are smaller than those for $A=1000$ [see Figs.~\ref{fig:DRIh2}(b3) and \ref{fig:DRIh1}(b3)], and $D_f$ decreases rapidly, as shown in Fig.~\ref{fig:DRIh1}(g). With further increase in $A$, this tendency of zigzag smearing and stochastic spike skipping in the non-stimulated sub-population is intensified, and eventually $D_f$ arrives at its minimum ($\simeq 0.548$) for $A=A_{min}^{(2)} (\simeq 4876)$. As an example near this minimum, we consider the case of $A=5000$. For this case, stochastic spike skipping is more intensified [see Fig.~\ref{fig:DRIh1}(d4)], and hence the original full synchronization in the non-stimulated sub-population becomes broken up (i.e., sparse stripes in the raster plot of spikes appear). Particularly, such sparse spiking stripes of non-stimulated FS interneurons are smeared in a zigzag way much more than those for the case of $A=3000$ [compare Fig.~\ref{fig:DRIh2}(a4) with Fig.~\ref{fig:DRIh2}(a3)]. Consequently, the amplitudes of $R_s^{(2)}(t)$ and $R_w(t)$ are much smaller than those for $A=3000$ [see Figs.~\ref{fig:DRIh2}(b4) and \ref{fig:DRIh1}(b4)] (i.e., the degree of population synchronization is reduced more significantly when compared with that for $A=3000$). As shown in the lower panel of Fig.~\ref{fig:DRIh2}(c4), peaks at the driving frequency $f_d$ and its harmonics also begin to be ``disrupted'' [i.e., their heights become smaller, and near $f_d$ new tiny peaks (of frequencies 164 and 183 Hz) appear]. However, with further increase in $A$ from $A_{min}^{(2)}$, non-stimulated FS interneurons begin to reorganize their spikings and exhibit a new type of sparse synchronization with the sub-population frequency $f_{sp}^{(2)} (\simeq 143$ Hz), along with enhanced external phase lockings of burstings of stimulated FS interneurons with the sub-population frequency $f_{sp}^{(1)} (\simeq 200$ Hz) [e.g., see the raster plots of spikes in Fig.~\ref{fig:DRIh2}(a5), the ISPSR kernel estimates $R_s^{(1)}(t)$ and $R_s^{(2)}(t)$ in Fig.~\ref{fig:DRIh2}(b5), and the power spectra in Fig.~\ref{fig:DRIh2}(c5) for $A=8000$].  A new peak, associated with sparse synchronization of non-stimulated FS interneurons, appears at $f \simeq 143$ Hz, as shown in the lower panel of Fig.~\ref{fig:DRIh2}(c5). (For $A=8000$, the peak at the driving frequency $f_d$ also coexists, but eventually it disappears for larger $A$ [see Figs.~\ref{fig:DRIh2}(c6)-\ref{fig:DRIh2}(c8)]). This ``sparse-synchronization'' peak of 143 Hz comes from evolution of the (above) tiny peak of 164 Hz for $A=5000$. With increasing $A$ the frequency of the tiny peak at 164 Hz for $A=5000$ becomes smaller, and for $A=8000$ the peak becomes broad and its frequency becomes 143 Hz. (On the other hand, as $A$ is increased the height of another peak of 183 Hz for $A=5000$ becomes smaller and it disappears.) Thanks to increase in the degree of synchronization in both the stimulated and the non-stimulated sub-populations, the amplitudes of both $R_s^{(1)}(t)$ and $R_s^{(2)}(t)$ become larger than those for $A=5000$ [compare Fig.~\ref{fig:DRIh2}(b5) with Fig.~\ref{fig:DRIh2}(b4)], which leads to the increase of the amplitude of $R_w(t)$ [see Fig.~\ref{fig:DRIh1}(b5)]. As a result, $D_f$ is increased, as shown in Fig.~\ref{fig:DRIh1}(g). As $A$ is further increased, external phase lockings of burstings with $f_{sp}^{(1)} \simeq 200$ Hz in the stimulated sub-population are more and more enhanced due to increased stimulation, while the degree of sparse synchronization in the non-stimulated sub-population becomes worse due to stochastic spike skipping of major non-stimulated FS interneurons and smearing of sparse stripes, as shown in the raster plots, the ISPSR kernel estimates $R_s^{(1)}(t)$ and $R_s^{(2)}(t)$, and the power spectra for $A=10^4$, $3 \times 10^4,$ and $6 \times 10^4$ [see Figs.~\ref{fig:DRIh2}(a6)-\ref{fig:DRIh2}(a8), Figs.~\ref{fig:DRIh2}(b6)-\ref{fig:DRIh2}(b8), and Figs.~\ref{fig:DRIh2}(c6)-\ref{fig:DRIh2}(c8)]; $f_{sp}^{(2)} \simeq$ 145, 146, and 146 Hz for $A=10^4$, $3 \times 10^4,$ and $6 \times 10^4$, respectively. Thanks to the dominance of external phase lockings in the stimulated sub-population, the overall degree of population synchronization in the whole population becomes better [i.e., the amplitudes of $R_w(t)$ increase, as shown in Figs.~\ref{fig:DRIh1}(b6)-\ref{fig:DRIh1}(b8)], and hence $D_f$ increases monotonically with $A$. Eventually when passing a threshold of $A^*_3 (\simeq 49699)$, $D_f$ becomes larger than 1, and then the 2nd stage appears where synchronization enhancement occurs, as shown in Fig.~\ref{fig:DRIh1}(g).

We also characterize the population synchronization in each of the stimulated and the non-stimulated sub-populations by employing the average occupation degree $\langle \langle O_i^{(l)} \rangle \rangle_r$, the average pacing degree $\langle \langle P_i^{(l)} \rangle \rangle_r$, and the average statistical-mechanical spiking measure $\langle M_s^{(l)} \rangle_r$; $l=1$ (2) represents the stimulated (non-stimulated) case. Plots of $\langle \langle O_i^{(l)} \rangle \rangle_r$, $\langle \langle P_i^{(l)} \rangle \rangle_r$, and $\langle M_s^{(l)} \rangle_r$ versus $A$ are given in Figs.~\ref{fig:DRIh2}(d1)-\ref{fig:DRIh2}(d3), respectively.
The average occupation degree $\langle \langle O_i^{(1)} \rangle \rangle_r$ is 1 (i.e., full synchronization occurs), independently of $A$, because every stimulated FS interneuron fires in each spiking stripe or bursting band.
This inhibitory full synchronization in the stimulated sub-population also exhibits high pacing degree $\langle \langle P_i^{(1)} \rangle \rangle_r$, similar to the excitatory case.
As $A$ is increased from 0, $\langle \langle P_i^{(1)} \rangle \rangle_r$ begins to decrease, and arrives at a minimum ($\simeq 0.737$) for $A \simeq 1320$. Near the minimum point, stimulated interneurons show mixed burstings and spikings, as shown in Fig.~\ref{fig:DRIh1}(a2) for $A=1000$ where each bursting consists of two spikes whose separation is wide. However, with further increase in $A$ external phase lockings of burstings of stimulated FS interneurons are more and more developed [see the developed bursting bands in Figs.~\ref{fig:DRIh2}(a3)-\ref{fig:DRIh2}(a8)]. Consequently, $\langle \langle P_i^{(1)} \rangle \rangle_r$ begins to increase, and it approaches a limit value ($ \simeq 0.82$). For this type of full synchronization (i.e., $\langle \langle O_i^{(1)} \rangle \rangle_r=1$), the average spiking measure $\langle M_s^{(1)} \rangle_r$ is the same as $\langle \langle P_i^{(1)} \rangle \rangle_r$. Unlike the stimulated case, when passing a threshold ($A \simeq 55$) (major) non-stimulated FS interneurons begin to exhibit stochastic spike skipping (i.e., intermittent and irregular spikings) due to a destructive effect of $S(t)$ (resulting from strong synaptic inhibition). Hence, $\langle \langle O_i^{(2)} \rangle \rangle_r$ varies depending on $A$ in the non-stimulated sub-population. Below the threshold $\langle \langle O_i^{(2)} \rangle \rangle_r =1$ (i.e., full synchronization takes place). However, above the threshold, sparse synchronization with $\langle \langle O_i^{(2)} \rangle \rangle_r <1$ occurs [i.e., sparse stripes appears in the raster plots of spikes. as shown in Figs.~\ref{fig:DRIh2}(a2)-\ref{fig:DRIh2}(a8)]. With increasing $A$ from the threshold, $\langle \langle O_i^{(2)} \rangle \rangle_r$ decreases monotonically, and its value becomes very low ($\simeq 0.23$) for large $A$, as shown in the lower panel of Fig.~\ref{fig:DRIh2}(d1), which is in contrast to the excitatory case of full synchronization [see in Fig.~\ref{fig:DREx2}(d1)]. As $A$ is increased from the threshold, zigzag smearing in the spiking stripes is more enhanced [see Figs.~\ref{fig:DRIh2}(a3)-\ref{fig:DRIh2}(a4)]. As a result, $\langle \langle P_i^{(2)} \rangle \rangle_r$ decreases rapidly, and it arrives at a minimum ($\simeq 0.265$) for $A \simeq 4981$, as shown in Fig.~\ref{fig:DRIh2}(d2). With increase in $A$ from the minimum point, such zigzag smearing begins to be reduced, and non-stimulated FS interneurons reorganize their spikings to exhibit a new type of sparse synchronization [compare Fig.~\ref{fig:DRIh2}(a5) with Fig.~\ref{fig:DRIh2}(a4)]. Then, $\langle \langle P_i^{(2)} \rangle \rangle_r$ increases a little, as shown in Fig.~\ref{fig:DRIh2}(d2). However, as $A$ is furtherer increased, sparse spiking stripes become more smeared [see Figs.~\ref{fig:DRIh2}(a6)-\ref{fig:DRIh2}(a8)], and hence $\langle \langle P_i^{(2)} \rangle \rangle_r$ decreases again; $\langle \langle P_i^{(2)} \rangle \rangle_r \simeq 0.26$ for large $A$.  For this case of sparse synchronization, the average spiking measure $\langle M_s^{(2)} \rangle_r$ is less than $\langle \langle P_i^{(2)} \rangle \rangle_r$ because $\langle \langle O_i^{(2)} \rangle \rangle_r < 1$, unlike the full synchronization which occurs in the stimulated case and in the excitatory case.

For examination of the matching degree between the stimulated and the non-stimulated sub-populations, we get the cross-correlation functions $\langle C_{12}(\tau) \rangle_r$ between $R_s^{(1)}(t)$ and $R_s^{(2)}(t)$ of the two sub-populations, which are shown for various values of $A$ in Figs.~\ref{fig:DRIh2}(e1)-\ref{fig:DRIh2}(e8). A plot of the cross-correlation measure $\langle M_c \rangle_r$ of Eq.~(\ref{eq:CM}) versus $A$ is also given in Fig.~\ref{fig:DRIh2}(f). Unlike the excitatory case, as $A$ is increased from 0 $\langle M_c \rangle_r$ decreases monotonically to its minimum ($\simeq -0.242$) for $A \simeq 4890$ (which is nearly the same as $A_{min}^{(2)} (\simeq 4876)$ for the minimum of $\langle D_f \rangle_r$) due to a destructive effect of external stimulus $S(t)$ (causing the zigzag smearing and the stochastic spike skipping in the non-stimulated sub-population). Because of monotonic decrease in $\langle M_c \rangle_r$, $\langle D_f \rangle_r$ also decreases from 1, and synchronization suppression occurs. After passing the minimum point ($A \simeq 4890$), $\langle M_c \rangle_r$ begins to increase slowly with $A$, but eventually it approaches 0 (without further increase), in contrast to the excitatory case (where $\langle M_c \rangle_r$ continue to increase monotonically without saturation) [compare Fig.~\ref{fig:DRIh2}(f) with Fig.~\ref{fig:DREx2}(f)]. We also note that the oscillating amplitudes of $\langle C_{12}(\tau) \rangle_r$ decrease with $A$, as shown in  Figs.~\ref{fig:DRIh2}(e5)-\ref{fig:DRIh2}(e8), unlike the excitatory case where the oscillating amplitudes of $\langle C_{12}(\tau) \rangle_r$ increase with $A$ [see Figs.~\ref{fig:DREx2}(e5)-\ref{fig:DREx2}(e8)]. This weak cross-correlation between the stimulated and the non-stimulated sub-populations occurs due to completely different types of population behaviors in the two sub-populations: non-stimulated FS interneurons exhibit sparse synchronization of low degree (without any external phase lockings), while stimulated FS interneurons show external phase lockings of burstings. Due to stronger stimulation effect, external phase lockings of stimulated FS interneurons are more and more intensified, and they become dominant. As a result, with increasing $A$ the overall degree of population synchronization in the whole population becomes better. Hence, both the amplitude of $R_w(t)$ and $\langle D_f \rangle_r$ increase monotonically with $A$ (without saturation) [see Figs.~\ref{fig:DRIh1}(b5)-\ref{fig:DRIh1}(b8) and Fig.~\ref{fig:DRIh1}(g)], in spite of weak cross-correlations between the two sub-populations. However, the increasing rate for $\langle D_f \rangle_r$ is much slower when compared with that for the excitatory case where the increase in $\langle D_f \rangle_r$ results from cooperation of the two sub-populations with strong cross-correlations.

\subsubsection{Large-Coupling Case of $J=1000$}
\label{subsubsec:LJ}

\begin{figure}
\includegraphics[width=0.8\columnwidth]{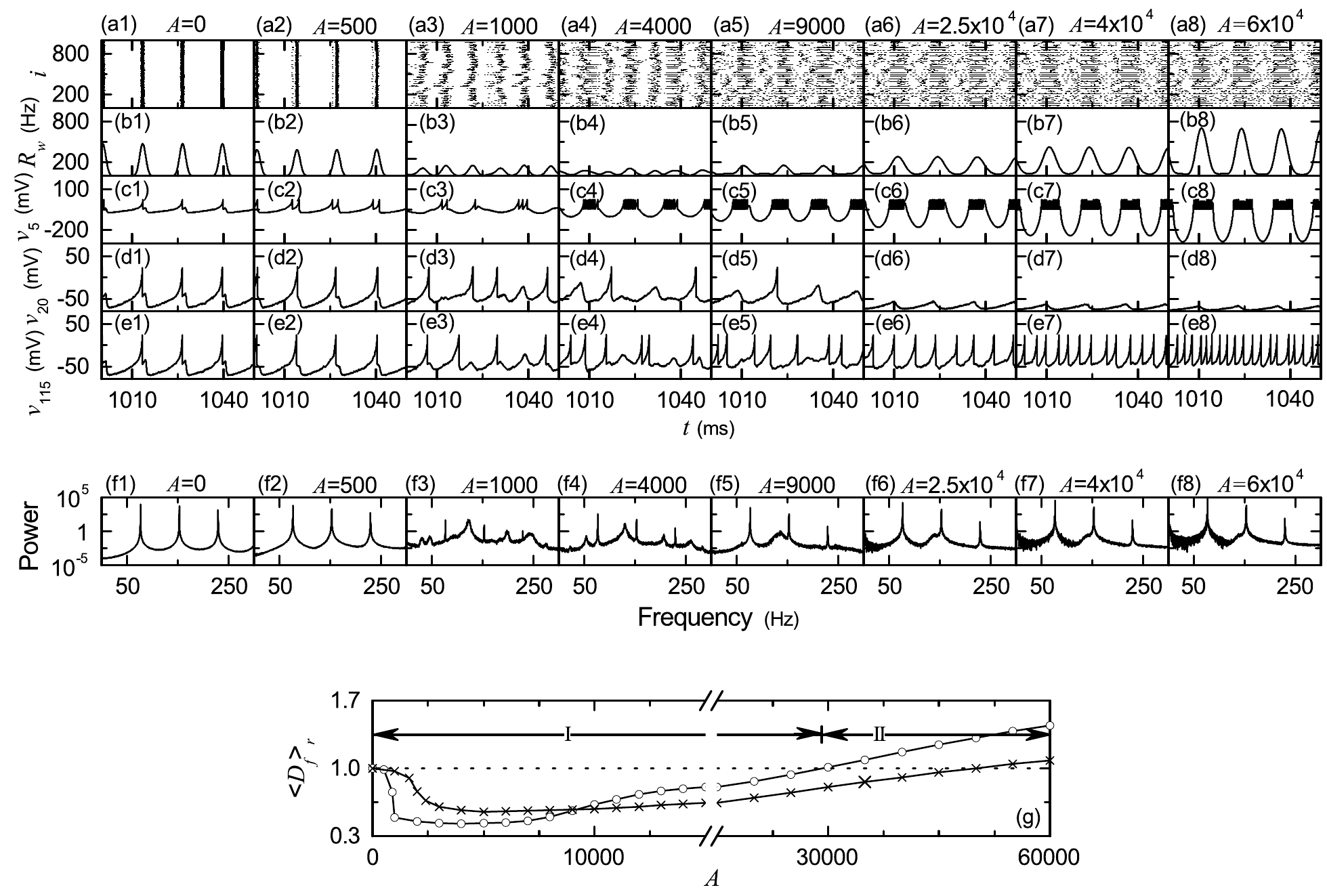}
\caption{Dynamical response when an external time-periodic stimulus $S(t)$ is applied to 50 randomly-chosen Izhikevich FS interneurons in the case of inhibitory synchronization for $J=1000$. We vary the driving amplitude $A$ for a fixed driving angular frequency $\omega_d$ (=0.48 rad/ms). Raster plots of spikes, instantaneous whole-population spike rate (IWPSR) kernel estimates $R_w(t)$, membrane potentials $v_5(t)$ of the stimulated 5th FS interneuron, membrane potentials $v_{20}(t)$ of the major non-stimulated 20th FS interneuron, and membrane potentials $v_{115}(t)$ of the minor non-stimulated 115th FS interneuron are shown for various values of $A$ in (a1)-(a8), (b1)-(b8), (c1)-(c8), (d1)-(d8), and (e1)-(e8), respectively. One-sided power spectra of $\Delta R_w(t) [=R_w(t) - \overline{R_w(t)}]$ (the overbar represents the time average) with mean-squared amplitude normalization are also given in (f1)-(f8). The band width of the Gaussian kernel estimate for each IWPSR $R_w(t)$ is 1 ms, and each power spectrum is obtained from $2^{16} (=65536)$ data points. (g) Plot of dynamical response factor $\langle D_f \rangle_r$ (open circles) versus $A$, where I and II represent the 1st (synchronization suppression) and the 2nd (synchronization enhancement) stages, respectively; for comparison, the dynamical response factor $\langle D_f \rangle_r$ (crosses) for $J=100$ is also shown. Here, $\langle \cdots \rangle_r$ represents an average over 30 realizations, and averaging time for $D_f$ in each realization is $3 \times 10^4$ ms.
}
\label{fig:DRIh3}
\end{figure}

We now consider a large-coupling case of $J=1000$ for comparison with the small-coupling case of $J=100$. We apply an external time-periodic AC stimulus $S(t)$ to $N_s(=50)$ randomly-selected Izhikevich FS interneurons by fixing the driving angular frequency as $\omega_d (=2 \pi f_d)$ =0.48 rad/ms ($f_d = \langle f_i \rangle$ =76 Hz), and study the dynamical response of inhibitory full synchronization by varying the driving amplitude $A$. Population synchronization for various values of $A$ may be well seen in the raster plots of spikes which are shown in Figs.~\ref{fig:DRIh3}(a1)-\ref{fig:DRIh3}(a8). The IWPSR kernel estimates $R_w(t),$ exhibiting oscillatory behaviors, are also shown in Figs.~\ref{fig:DRIh3}(b1)-\ref{fig:DRIh3}(b8), and the power spectra of $\Delta R_w(t)$ are given in Figs.~\ref{fig:DRIh3}(f1)-\ref{fig:DRIh3}(f8). Moreover, times series of membrane potentials of individual FS interneurons are given for various values of $A$. The time series of $v_5(t)$ of the 5th stimulated FS interneuron are shown in Figs.~\ref{fig:DRIh3}(c1)-\ref{fig:DRIh3}(c8). As explained in the case of $J=100$, there are two types of non-stimulated FS interneurons, depending on their synaptic connections. Major non-stimulated FS interneurons (which have synaptic connections with fast-firing stimulated FS interneurons) fire slowly due to increased inhibition, while minor non-stimulated FS interneurons (which have no direct synaptic connections with stimulated FS interneurons and receive synaptic inputs from major slowly-firing non-stimulated FS interneurons) fire fast spikings due to decreased inhibition. Figures \ref{fig:DRIh3}(d1)-\ref{fig:DRIh3}(d8) show the time series of $v_{20}(t)$ of the 20th major slowly-firing non-stimulated FS interneuron. On the other hand,  Figs.~\ref{fig:DRIh3}(e1)-\ref{fig:DRIh3}(e8) show the time series of $v_{115}(t)$ of the 115th minor fast-firing non-stimulated FS interneuron. A plot of the dynamical factor $\langle D_f \rangle_r$ versus $A$ is shown in Fig.~\ref{fig:DRIh3}(g). Like the case of $J=100$, two stages are thus found to appear. Synchronization suppression ($\langle D_f \rangle_r <1$) and synchronization enhancement ($\langle D_f \rangle_r >1$) occur in the 1st (I) stage ($0< A < A^*_4$) and the 2nd (II) stage ($A > A^*_4$), respectively, where $A^*_4 \simeq 29207$ [which is less than $A^*_3 (\simeq 49699)$ for the case of $J=100$]. Examples are given for various values of $A$; 1st stage ($A=500$, 1000, 4000, 9000, and $2.5 \times 10^4$), and 2nd stage ($A= 4.0 \times 10^4,$ and $6 \times 10^4$).

\begin{figure}
\includegraphics[width=0.8\columnwidth]{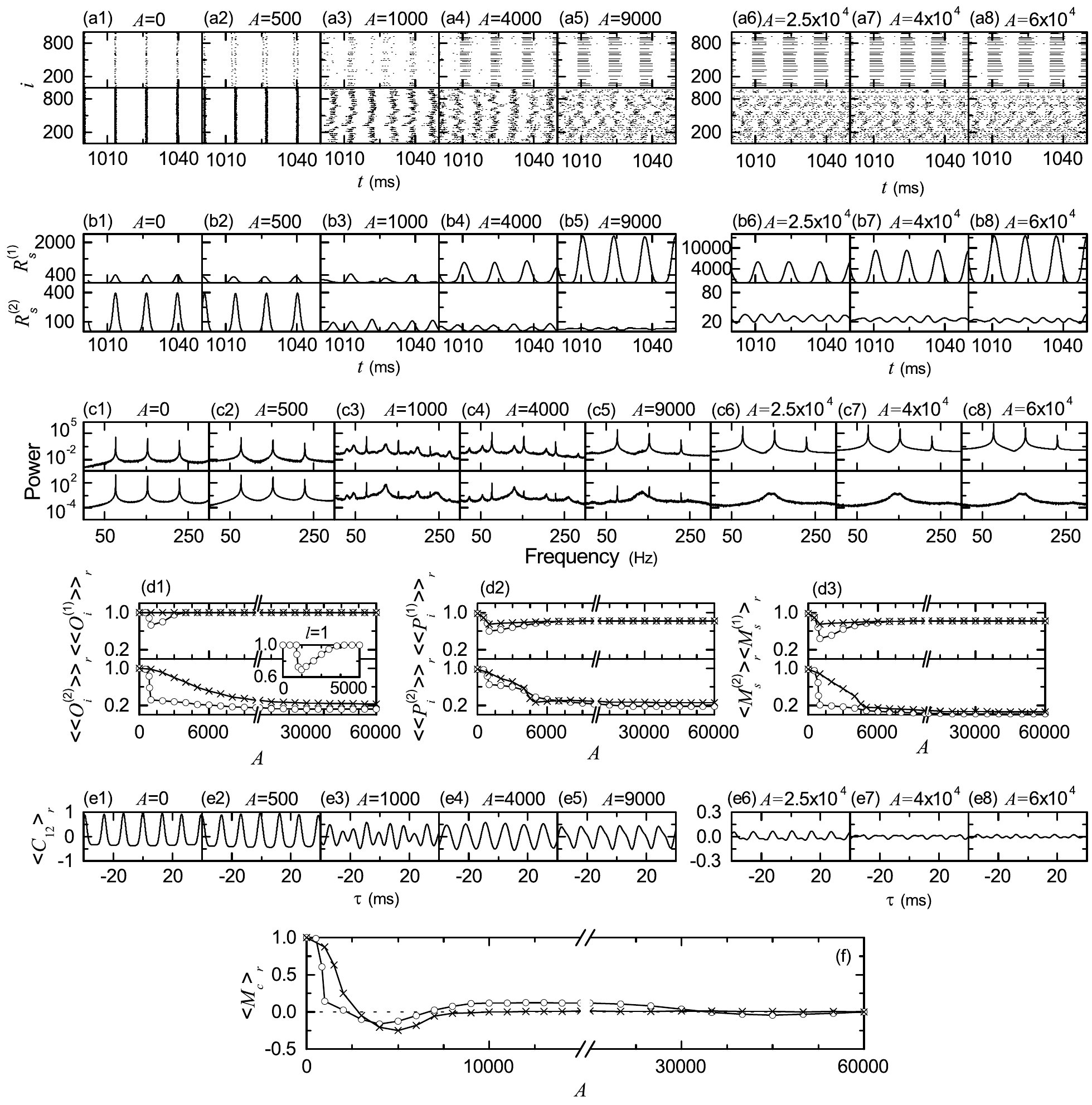}
\linespread{0.8}
\caption{Dynamical responses for the case of Fig.~\ref{fig:DRIh3} in both the stimulated and the non-stimulated sub-populations and a cross-correlation measure $\langle M_c \rangle_r$ between the dynamics of the two sub-populations. Raster plot of spikes, instantaneous sub-population spike rates (ISPSRs) $R_s^{(1)}$ and $R_s^{(2)}$ [the superscript 1 (2) corresponds to the stimulated (non-stimulated) case], and one-sided power spectra of $\Delta R_s^{(1)}(t) [=R_s^{(1)}(t) - \overline{R_s^{(1)}(t)}]$ and  $\Delta R_s^{(2)}(t) [=R_s^{(2)}(t) - \overline{R_s^{(2)}(t)}]$ (the overbar represents the time average) with mean-squared amplitude normalization in the stimulated and the non-stimulated sub-populations are shown for various values of $A$ in (a1)-(a8), (b1)-(b8), and (c1)-(c8), respectively: the upper (lower) panels denote those for the stimulated (non-stimulated) case. The band width of the Gaussian kernel estimate for each ISPSR is 1 ms, and each power spectrum is obtained via 30 realizations [$2^{16} (=65536)$ data points are used in each realization]. Plots of the average occupation degree $\langle \langle O_i^{(l)} \rangle \rangle_r$, the average pacing degree $\langle \langle P_i^{(l)} \rangle \rangle_r$, and the statistical-mechanical spiking measure $\langle M_s^{(l)} \rangle_r$ (denoted by open circles) versus $A$ are shown in (d1)-(d3), respectively; $l=1$ (2) corresponds to the stimulated (non-stimulated) case. Here, $\langle O_i^{(l)} \rangle$, $\langle P_i^{(l)} \rangle$, and $M_s^{(l)}$ are obtained by following the $3 \times 10^3$ stripes/bands in the raster plot of spikes in each realization, and $\langle \cdots \rangle_r$ denotes an average over 30 realizations; for comparison, those for $J=100$ are represented by crosses. Plots of cross-correlation functions $\langle C_{12}(\tau) \rangle_r$ versus $\tau$ are also given in (e1)-(e8). The number of data used for the calculation of each temporal cross-correlation function $C_{12}(\tau)$ is $2^{16} (=65536)$ in each realization, and $\langle C_{12}(\tau) \rangle_r$ is obtained through an average over 30 realizations. (f) Plot of the cross-correlation measure $\langle M_c \rangle_r$ (open circles) versus $A$; $\langle M_c \rangle_r$ is obtained via an average over 30 realizations. For comparison, $\langle M_c \rangle_r$ (crosses) for $J=100$ is also shown.
}
\label{fig:DRIh4}
\end{figure}

As in the above case of $J=100$, we make a detailed analysis of dynamical responses by decomposing the whole population of FS interneurons into two sub-populations of the stimulated and the non-stimulated FS interneurons. Dynamical responses in these two sub-populations are given in Fig.~\ref{fig:DRIh4}. Raster plots of spikes, ISPSR kernel estimates $R_s^{(1)}(t)$ and $R_s^{(2)}(t)$ [the superscript 1 (2) corresponds to the stimulated (non-stimulated) case], and power spectra of $\Delta R_s^{(1)}(t)$ and $\Delta R_s^{(2)}(t)$ in the stimulated and the non-stimulated sub-populations are shown in Figs.~\ref{fig:DRIh4}(a1)-\ref{fig:DRIh4}(a8), Figs.~\ref{fig:DRIh4}(b1)-\ref{fig:DRIh4}(b8), and Figs.~\ref{fig:DRIh4}(c1)-\ref{fig:DRIh4}(c8), respectively: the upper (lower) panels in these figures denote those for the stimulated (non-stimulated) case.
For characterization of population synchronization in each of the stimulated and the non-stimulated sub-populations, the average occupation degree $\langle \langle O_i^{(l)} \rangle \rangle_r$, the average pacing degree $\langle \langle P_i^{(l)} \rangle \rangle_r$, and the average statistical-mechanical spiking measure $\langle M_s^{(l)} \rangle_r$ are given in Figs.~\ref{fig:DRIh4}(d1)-\ref{fig:DRIh4}(d3), respectively; $l=1$ (2) represents the stimulated (non-stimulated) case. The cross-correlation functions $C_{12}(\tau)$ between $R_s^{(1)}(t)$ and $R_s^{(2)}(t)$ of the two sub-populations are shown for various values of $A$ in Figs.~\ref{fig:DRIh4}(e1)-\ref{fig:DRIh4}(e8). Figure \ref{fig:DRIh4}(f) shows a plot of the cross-correlation measure $\langle M_c \rangle_r$ of Eq.~(\ref{eq:CM}) versus $A$.

As $A$ is increased from 0 and passes a threshold, stimulated FS interneurons begin to exhibit burstings, as shown in Fig.~\ref{fig:DRIh3}(c2) for $A=500$. These burstings are phase-locked to external stimulus $S(t)$. For this case, spikings of non-stimulated FS interneurons cannot be well matched with burstings of stimulated FS interneurons, because of difference in the type of firings of individual neurons [e.g., see the raster plots of spikes in Figs.~\ref{fig:DRIh3}(a2) and \ref{fig:DRIh4}(a2) for $A=500$]. However, these spiking stripes of non-stimulated FS interneurons are also phase-locked to external stimulus, although they are phase-shifted from the vertical bursting trains of stimulated FS interneurons. Peaks in the power spectra of $\Delta R_s^{(1)}(t)$ and $\Delta R_s^{(2)}(t)$, associated with external phase lockings for both cases of stimulated and non-stimulated FS interneurons, appear at the driving frequency $f_d $ (=76 Hz) and its harmonics [see Fig.~\ref{fig:DRIh4}(c2)].

As $A$ is further increased and passes another threshold $(\simeq 894$), single-periodic synchronization disappears and a new type of multi-periodic synchronization occurs abruptly for both cases of the stimulated and the non-stimulated sub-populations in a wide region of $A$, in contrast to the above case of $J=100$ where multi-periodic synchronization occurs only in the non-stimulated sub-population [e.g., see the lower panel of Fig.~\ref{fig:DRIh2}(c4) for $A=5000$]: this multi-periodicity for $J=1000$ ends earlier for the stimulated case ($A \sim 8900$) when compared with the non-stimulated case ($A \sim 23200$). In this intermediate range of $A$, major non-stimulated FS interneurons exhibit intermittent and stochastic spikings (i.e., stochastic spike skipping) [see Fig.~\ref{fig:DRIh3}(d3) for $A=1000$]. Due to stronger inhibition for $J=1000$, stochastic spike skipping occurs for smaller values of $A$ than those for $J=100$. Moreover, stimulated FS interneurons also show intermittent and stochastic mixed burstings and spikings due to strong stochastic synaptic inputs from non-stimulated FS interneurons, as shown in Fig.~\ref{fig:DRIh3}(c3) for $A=1000$, in contrast to the case of $J=100$ where only regular burstings of stimulated FS interneurons become gradually intensified (i.e., for $J=100$ only single-periodic full synchronization of burstings occurs in the stimulated sub-population). Thus, for $A=1000$ multi-periodic synchronization with two fundamental frequencies (of 76 Hz and 123 Hz) appears, as shown in the power spectra of $\Delta R_s^{(1)}(t)$ and $\Delta R_s^{(2)}(t)$ in Fig.~\ref{fig:DRIh4}(c3) where peaks appear at the two fundamental frequencies, their harmonics, their sum (i.e., 199 Hz), and so on. Due to stochastic spike/burst skipping, sparse stripes appear in the raster plots of spikes in both the stimulated and the non-stimulated sub-populations [see Fig.~\ref{fig:DRIh4}(a3)], in contrast to the case of $J=100$ where sparse spiking stripes appear only in the non-stimulated case. Furthermore, zigzag smearing also occurs in the sparse spiking stripes for the case of the non-stimulated sub-population, as in the case of $J=100$, due to the high clustering coefficient of the Watts-Strogatz SWN. As a result, the overall degree of population synchronization in the whole population for $A=1000$ is much more reduced when compared with the case of $A=500$ [compare Fig.~\ref{fig:DRIh3}(b3) with Fig.~\ref{fig:DRIh3}(b2)]. Hence, the dynamical factor $\langle D_f \rangle_r$ is abruptly decreased until about $A=1000$, in comparison with the case of $J=100$, and then it arrives slowly at its minimum ($\simeq 0.413$) for $A=A_{min}^{(3)} (\simeq 3753)$ (which is smaller than $A=A_{min}^{(2)} (\simeq 4876)$ for $J=100$) [see Fig.~\ref{fig:DRIh3}(g)]; near the minima of $\langle D_f \rangle_r$ for both cases of $J=100$ and 1000, $\langle D_f \rangle_r$ for $J=1000$ is lower than that for $J=100$.

With further increase in $A$, the degree of stochastic skippings of stimulated FS interneurons is decreased, and they begin to exhibit more regular burstings. As a result of enhanced external phase lockings, distinct bursting bands appear in the raster plot of spikes, as shown in the upper panel of Fig.~\ref{fig:DRIh4}(a4) for $A=4000$. Peaks at the driving frequency $f_d$ (=76 Hz) and its harmonics, associated with external phase lockings, become sharper (i.e., their heights increase) than those for $A=1000$ in the power spectrum of $\Delta R_s^{(1)}(t)$ [compare the upper panel of Fig.~\ref{fig:DRIh4}(c4) with the upper panel of Fig.~\ref{fig:DRIh4}(c3)]. Hence, the amplitude of $R_s^{(1)}(t)$ is also larger than that for $A=1000$, as compared in Fig.~\ref{fig:DRIh4}(b4) and Fig.~\ref{fig:DRIh4}(b3). On the other hand, major non-stimulated FS interneurons show more stochastic spike skippings [see Fig.~\ref{fig:DRIh3}(d4)]. Moreover, sparse spiking stripes of non-stimulated FS interneurons for $A=4000$ are smeared in a zigzag way much more than those for $A=1000$, as shown in the lower panel of Fig.~\ref{fig:DRIh4}(a4), and hence the amplitude of $R_s^{(2)}(t)$ becomes smaller than that for $A=1000$ [compare Fig.~\ref{fig:DRIh4}(b4) with Fig.~\ref{fig:DRIh4}(b3)], in contrast to the stimulated case. For this case, the peak at another fundamental frequency $f (\simeq 130$ Hz) becomes sharper than that ($f \simeq 123$ Hz) for $A=1000$ in the power spectrum of $\Delta R_s^{(2)}(t)$ [compare the lower panel of Fig.~\ref{fig:DRIh4}(c4) with the lower panel of Fig.~\ref{fig:DRIh4}(c3)]. The overall degree of population synchronization for $A=4000$ is a little lower than that for $A=1000,$ mainly due to increased zigzag smearing in the non-stimulated sub-population [compare Fig.~\ref{fig:DRIh3}(b4) with Fig.~\ref{fig:DRIh3}(b3)].

However, as $A$ is further increased, stimulated FS interneurons begin to show single-periodic behavior (with only one fundamental frequency), associated with external phase lockings of burstings, as shown in the raster plot of spikes and the power spectrum of $\Delta R_s^{(1)}(t)$ in the upper panels of Figs.~\ref{fig:DRIh4}(a5) and \ref{fig:DRIh4}(c5) for $A=9000$. For this case, only peaks at the driving frequency $f_d (= 76$ Hz) and its harmonics, related to external phase lockings, persist (i.e., all the other old peaks disappear) in the power spectrum, and they become sharper. As a result of enhanced external phase locking of burstings, the amplitude of $R_s^{(1)}(t)$ is much increased, as shown in Fig.~\ref{fig:DRIh4}(b5). On the other hand, non-stimulated FS interneurons continue to exhibit multi-periodic behavior for $A=9000$. For this case, stochastic spike skipping and smearing are more enhanced, as shown in the lower panel of Fig.~\ref{fig:DRIh4}(a5). Consequently, the amplitude of $R_s^{(2)}(t)$ is reduced [see Fig.~\ref{fig:DRIh4}(b5)]. For this non-stimulated case, peaks in the power spectrum of $\Delta R_s^{(2)}(t)$ appear at two fundamental frequencies of 76 Hz and 137 Hz and their harmonics, as shown in the lower panel of Fig.~\ref{fig:DRIh4}(c5), in contrast to the case of stimulated sub-population. For this case of $A=9000$, the enhanced external phase lockings of burstings in the stimulated sub-population becomes dominant, and hence the overall degree of population synchronization for $A=9000$ is increased [see the increased amplitude of $R_w(t)$ in Fig.~\ref{fig:DRIh3}(b5)]. In this way, with increasing $A$ from $A_{min}^{(3)}$ the dynamical factor $\langle D_f \rangle_r$ increases gradually thanks to enhancement of external phase lockings of burstings of stimulated FS interneurons, as shown in Fig.~\ref{fig:DRIh3}(g). The increasing rate of $\langle D_f \rangle_r$ for $J=1000$ is larger than that for $J=100$ because stimulated FS interneurons receive weaker synaptic inputs from non-stimulated FS interneurons (resulting from more developed stochastic spike skipping of major non-stimulated interneurons). Eventually, as $A$ passes a threshold $(A \simeq 9183)$, $\langle D_f \rangle_r$ for $J=1000$ becomes larger than that for $J=100$ [see Fig.~\ref{fig:DRIh3}(g)]. When $A$ is sufficiently large, non-stimulated FS interneurons also begin to show single-periodic behaviors (with one fundamental frequency), as shown in the power spectrum of $\Delta R_s^{(2)}(t)$ for $A=25000$ [see the lower panel of Fig.~\ref{fig:DRIh4}(c6)] where only one fundamental frequency at $f \simeq 143$ Hz exists (i.e., all the other peaks, associated with external phase lockings, disappear). Then, similar to the case of $J=100$, stimulated FS interneurons exhibit regular bursting behaviors [with the sub-population frequency $f_{sp}^{(1)} (\simeq 76$ Hz)], while non-stimulated FS interneurons show fast sparse synchronization [with the sub-population frequency $f_{sp}^{(2)} (\simeq 143$ Hz)] [see the raster plots of spikes in Fig.~\ref{fig:DRIh4}(a6), the ISPSR kernel estimates $R_s^{(1)}(t)$ and $R_s^{(2)}(t)$ in Fig.~\ref{fig:DRIh4}(b6), and the power spectra in Fig.~\ref{fig:DRIh4}(c6)]. With increasing $A$ furthermore, external phase lockings of burstings with $f_{sp}^{(1)} \simeq 76$ Hz in the stimulated sub-population become more and more enhanced than those for $J=100$, thanks to both increased external stimulation and weaker synaptic inputs from non-stimulated FS interneurons, as shown in the raster plots of spikes, the ISPSR kernel estimate $R_s^{(1)}(t)$, and the power spectra for $A= 4 \times 10^4$ and $6 \times 10^4$ [see the upper panels of Figs.~\ref{fig:DRIh4}(a7)-\ref{fig:DRIh4}(a8), Figs.~\ref{fig:DRIh4}(b7)-\ref{fig:DRIh4}(b8), and Figs.~\ref{fig:DRIh4}(c7)-\ref{fig:DRIh4}(c8)]. On the other hand, the degree of fast sparse synchronization in the non-stimulated sub-population becomes very low due to stochastic spike skipping and smearing, as shown in the raster plots and the ISPSR kernel estimate $R_s^{(2)}(t)$ for $A= 4 \times 10^4$ and $6 \times 10^4$ [see the lower panels of Figs.~\ref{fig:DRIh4}(a7)-\ref{fig:DRIh4}(a8) and Figs.~\ref{fig:DRIh4}(b7)-\ref{fig:DRIh4}(b8)];
$f_{sp}^{(2)} \simeq 146$ Hz for $A= 4 \times 10^4$ and $6 \times 10^4$, as shown in the lower panels of Figs.~\ref{fig:DRIh4}(c7)-\ref{fig:DRIh4}(c8)]. Thanks to the dominance of more-developed external phase lockings of burstings of stimulated FS interneurons, $\langle D_f \rangle_r$ increases with $A$ in a faster rate than that for $J=100$, as shown in Fig.~\ref{fig:DRIh3}(g). Eventually when passing a threshold of $A^*_4 (\simeq 29207)$ [which is smaller than $A^*_3 (\simeq 49699)$ for $J=100$], $\langle D_f \rangle_r$ becomes larger than 1, and then the 2nd stage of synchronization enhancement occurs.

For characterization of the population synchronization in each of the stimulated and the non-stimulated sub-populations, we employ the average occupation degree $\langle \langle O_i^{(l)} \rangle \rangle_r$, the average pacing degree $\langle \langle P_i^{(l)} \rangle \rangle_r$, and the average statistical-mechanical spiking measure $\langle M_s^{(l)} \rangle_r$; $l=1$ (2) denotes the stimulated (non-stimulated) case. Plots of $\langle \langle O_i^{(l)} \rangle \rangle_r$, $\langle \langle P_i^{(l)} \rangle \rangle_r$, and $\langle M_s^{(l)} \rangle_r$ versus $A$ are shown in Figs.~\ref{fig:DRIh4}(d1)-\ref{fig:DRIh4}(d3), respectively. For small $A$, stimulated FS interneurons exhibit regular spikings or burstings, which results in the full synchronization with $\langle \langle O_i^{(1)} \rangle \rangle_r =1$. However, when passing a threshold ($A \simeq 894$), stimulated  FS interneurons exhibit stochastic spike/burst skippings due to strong stochastic synaptic inputs from non-stimulated FS interneurons [e.g., see Fig.~\ref{fig:DRIh3}(c3) for $A=1000$], in contrast to the case of $J=100$ where stimulated FS interneurons exhibit only regular burstings/spikings without skippings. Due to this stochastic skipping, $\langle \langle O_i^{(1)} \rangle \rangle_r$ becomes less than 1 (i.e. sparse synchronization appears) [see the inset of Fig.~\ref{fig:DRIh4}(d1)], unlike the case of full synchronization for $J=100$. However, as $A$ is further increased, external phase lockings of burstings of stimulated FS interneurons are more and more enhanced, and full synchronization with $\langle \langle O_i^{(1)} \rangle \rangle_r =1$ reappears when passing a higher threshold ($A \simeq 4159$). This stochastic spike/burst skipping of stimulated FS interneurons also affects the average pacing degree $\langle \langle P_i^{(1)} \rangle \rangle_r$. As $A$ is increased from 0, $\langle \langle P_i^{(1)} \rangle \rangle_r$ begins to decrease and arrives at its minimum ($ \simeq 0.532$) for $A \simeq 1208$ due to the stochastic skippings. Near this minimum point, $\langle \langle P_i^{(1)} \rangle \rangle_r$ is less than that for $J=100$, as shown in Fig.~\ref{fig:DRIh4}(d2). With further increase in $A$, $\langle \langle P_i^{(1)} \rangle \rangle_r$ begins to increase, and it approaches a limit value ($\simeq 0.82$) (which seems to be the same as that for $J=100$) thanks to enhancement of external phase lockings of burstings of stimulated FS interneurons. Consequently, $\langle M_s^{(1)} \rangle_r$ near the minimum point is much less than that for $J=100$ due to sparse synchronization (with $\langle \langle O_i^{(1)} \rangle \rangle_r < 1$), while for large $A$ the values of $\langle M_s^{(1)} \rangle_r$ for both cases of $J=1000$ and 100 seem to be the same thanks to developed external phase lockings of burstings of stimulated FS interneurons.
When passing the threshold ($A \simeq 894$), (major) non-stimulated FS interneurons also begin to exhibit stochastic spike skippings due to strong inhibition from stimulated FS interneurons. The degree of stochastic skippings is more severe than that for $J=100$ since the strength of synaptic inhibition is stronger. As a result, the original full synchronization breaks up, and fast sparse synchronization occurs, as shown in the lower panel of Fig.~\ref{fig:DRIh4}(a3) for $A=1000$ [where more sparse spiking stripes appear due to fast oscillation with the sub-population frequency $f_s^{(2)} (\simeq 123$ Hz)]. For this case, $\langle \langle O_i^{(2)} \rangle \rangle_r $ (i.e., the average occupation degree in each spiking stripe) is decreased more abruptly when compared with the case of $J=100$ [see Fig.~\ref{fig:DRIh4}(d1)]. Thus, $\langle \langle O_i^{(2)} \rangle \rangle_r$ becomes much less than that for $J=100$. After that, $\langle \langle O_i^{(2)} \rangle \rangle_r$ decreases slowly, but it is still less than that for $J=100$. With increasing $A$ from the threshold, zigzag smearing in sparse spiking stripes is more developed [see Figs.~\ref{fig:DRIh4}(a3)-\ref{fig:DRIh4}(a4)]. As a result, $\langle \langle P_i^{(2)} \rangle \rangle_r$ is decreased in a relatively rapid way, as shown in Fig.~\ref{fig:DRIh4}(d2). As $A$ is further increased, sparse spiking stripes become smeared gradually [see Figs.~\ref{fig:DRIh4}(a5)-\ref{fig:DRIh4}(a8)], which also leads to gradual decrease in $\langle \langle P_i^{(2)} \rangle \rangle_r$. For large $A,$ $\langle \langle P_i^{(2)} \rangle \rangle_r \simeq 0.18$, which is less than that for $J=100$. Like the case of $\langle \langle O_i^{(2)} \rangle \rangle_r,$ the average spiking measure $\langle M_s^{(2)} \rangle_r$ also shows an abrupt decrease when fast sparse synchronization appears, and then it decreases slowly [see Fig.~\ref{fig:DRIh4}(d3)]. Due to stronger destructive effect of $S(t)$ (resulting from strong synaptic inhibition), the value of $\langle M_s^{(2)} \rangle_r$ becomes less than that for $J=100$.

Finally, we obtain the cross-correlation functions $\langle C_{12}(\tau) \rangle_r$ between $R_s^{(1)}(t)$ and $R_s^{(2)}(t)$ of the two sub-populations, which are shown for various values of $A$ in Figs.~\ref{fig:DRIh4}(e1)-\ref{fig:DRIh4}(e8), and examine the matching degree between the stimulated and the non-stimulated sub-populations. A plot of the cross-correlation measure $\langle M_c \rangle_r$ of Eq.~(\ref{eq:CM}) versus $A$ is also shown in Fig.~\ref{fig:DRIh4}(f): $\langle M_c \rangle_r$ for $J=100$ is also given for comparison. In contrast to the case of $J=100$, $\langle M_c \rangle_r$ decreases abruptly until about $A=1000$, mainly due to stochastic skipping of firings in both the stimulated and the non-stimulated sub-populations, and then it arrives at its minimum ($\simeq -0.13$) in a relatively slow way for $A \simeq 3803$ (which is  nearly the same as $A_{min}^{(3)} (\simeq 3753)$ for the minimum of $\langle D_f \rangle_r$). Because of the abrupt decrease in $\langle M_c \rangle_r$, the dynamical factor $\langle D_f \rangle_r$ also decreases rapidly from 1, and synchronization suppression occurs. After passing the minimum point ($A \simeq 3803$), $\langle M_c \rangle_r$ begins to increase slowly with $A$, but eventually it approaches 0 in an oscillatory way [see Fig.~\ref{fig:DRIh4}(f)].
In addition to $\langle M_c \rangle_r$ which is given by $\langle C_{12}(0) \rangle_r$ at the zero-time lag, we are also concerned about the peak amplitude (i.e., maximum amplitude) $\langle C_{12}(\tau_{peak}) \rangle_r$  at the ``peak'' time lag $\tau_{peak}$. For large $A$, the peak amplitude decreases with $A$ much more rapidly than that for $J=100$ [compare Figs.~\ref{fig:DRIh4}(e6)-\ref{fig:DRIh4}(e8) with Figs.~\ref{fig:DRIh2}(e6)-\ref{fig:DRIh2}(e8)]. Consequently, the cross-correlation between the stimulated and the non-stimulated sub-populations becomes very weak because of distinctly different types of population behaviors in the two sub-populations. Stimulated FS interneurons exhibit external phase lockings of burstings, while non-stimulated FS interneurons show fast sparse synchronization (without any external phase lockings). With increasing $A$, external phase lockings of stimulated FS interneurons become more and more intensified thanks to stronger stimulation and weaker synaptic inputs, while the degree of sparse synchronization in the non-stimulated sub-population is negligibly low. As a result of dominant effect of such external phase lockings, the dynamical factor $\langle D_f \rangle_r$ increases monotonically with $A$ [see Fig.~\ref{fig:DRIh3}(g)], in spite of weak cross-correlations between the two sub-populations, as in the case of $J=100$: the increasing rate for $J=1000$ is larger than that for $J=100$ due to more-developed external phase lockings. However, the increasing rates for $\langle D_f \rangle_r$ in both inhibitory cases of $J=100$ and 1000 are much slower than that for the excitatory case where the increase in $\langle D_f \rangle_r$ results from the interplay between the two sub-populations with strong cross-correlations.

\section{Summary}
\label{sec:SUM}
Brain rhythms appear in health and diseases via neural synchronization. A neural system's response to external stimulus can provide useful information about its dynamical properties. Therefore, it is important to investigate how an external stimulus affects the neural synchronization. Synchronization enhancement or suppression may occur via control of population synchronization. In most previous theoretical and computational works on control of population synchronization, only excitatory-type couplings were considered. To see the dependence of dynamical responses to external stimuli on the synaptic-coupling type, we considered two types of excitatory and inhibitory full synchronization in the Watts-Strogatz SWN of excitatory RS pyramidal neurons and inhibitory FS interneurons, and investigated the effects of synaptic interactions on dynamical responses to external time-periodic stimuli $S(t)$ by varying the driving amplitude $A$. We have characterized dynamical responses to $S(t)$ in terms of the dynamical response factor $\langle D_f \rangle_r$  by increasing $A$. For the case of excitatory coupling, external phase lockings occur in both the stimulated and the non-stimulated sub-populations, thanks to a constructive effect of $S(t)$ which results from phase-attractive synaptic excitation. On the other hand, in the case of inhibitory coupling, external phase locking occurs only in the stimulated sub-population, while the original inhibitory full synchronization in the non-stimulated sub-population breaks up gradually (i.e., for large $A$ non-stimulated FS interneurons exhibit inhibitory sparse synchronization of low degree) due to a destructive effect of $S(t)$ which comes from strong synaptic inhibition. As results of these different effects of $S(t)$, the type and degree of dynamical response (e.g., synchronization enhancement or suppression characterized by $\langle D_f \rangle_r$ ) have been found to vary differently, depending on the type of synaptic interaction. For a detailed analysis, we have also measured the matching degree between the dynamics of the two sub-populations of stimulated and non-stimulated neurons in terms of a cross-correlation measure $\langle M_c \rangle_r$. $\langle M_c \rangle_r$ has been found to vary with $A$ in a different way, depending on the synaptic-coupling type. For small $A$, synchronization enhancement occurs for the excitatory case, thanks to strong cross-correlation (with $\langle M_c \rangle_r > 0.97$) between the two sub-populations, while synchronization suppression takes place in the inhibitory case, due to monotonic decrease in $\langle M_c \rangle_r$. Particularly, for large $A$ the cross-correlation becomes very weak in the inhibitory case, while for the excitatory case $\langle M_c \rangle_r$ increases gradually after passing its minimum (i.e., it becomes large for large $A$). Consequently, in the excitatory case synchronization enhancement reappears for an intermediate value of $A$, thanks to the strong cross-correlation: with increasing $A$ from 0, synchronization enhancement first appears, then synchronization suppression occurs, and finally synchronization enhancement reappears. For the inhibitory case, in spite of weak cross-correlation, synchronization enhancement also appears for sufficiently large $A$, just thanks to much-enhanced external phase lockings of burstings of stimulated FS interneurons: with increase in $A$ from 0, synchronization suppression appears in a wide range of $A$, and then synchronization enhancement occurs for very large $A$. Furthermore, in the inhibitory case we have also studied the effect of coupling strength $J$ on the dynamical responses by considering both the small- and the large-coupling cases (i.e., $J=100$ and 1000). Thus, dynamical response has been found to vary in a quantitatively different way, depending on the coupling strength $J$. For intermediate values of $A$ stimulated FS interneurons in the case of $J=1000$ have been found to exhibit intermittent and stochastic mixed burstings and spikings due to strong stochastic synaptic inputs from non-stimulated FS interneurons, in contrast to the case of $J=100$ where they show only regular burstings. As a result, the dynamical factor $\langle D_f \rangle_r$ decreases rapidly in comparison with the case of $J=100$. However, after passing its minimum, $\langle D_f \rangle_r$  increases faster than that for $J=100$, and eventually when passing an intermediate threshold it becomes larger, thanks to much more enhanced external phase lockings of burstings of stimulated FS interneurons (resulting from both stronger external stimulation and weaker synaptic inputs from non-stimulated FS interneurons). All these results for both excitatory and inhibitory cases are expected to provide useful insights on the dynamical responses to external stimuli in neural systems (i.e., how external stimuli affect brain rhythms emerging via excitatory and inhibitory synchronization).

\begin{acknowledgments}
This research was supported by Basic Science Research Program through the National Research Foundation of Korea (NRF) funded by the Ministry of Education
(Grant No. 20162007688).
\end{acknowledgments}

\appendix

\section{Statistical-Mechanical Spiking Measure in The Stimulated and The Non-stimulated Sub-Populations}
\label{sec:SMSM}
We measure the degree of population synchronization in each of the stimulated and the non-stimulated sub-populations in terms of a realistic statistical-mechanical spiking measure, based on the ISPSR kernel estimate
$R_s^{(l)}(t)$ ($l=1$ and 2 correspond to the stimulated and the non-stimulated cases, respectively) \cite{Kim1}. Population synchronization may be well visualized in the raster plot of spikes. For a synchronized case, spiking stripes or bursting bands (indicating population synchronization) appear successively in the raster plot, and the corresponding ISPSR kernel estimate $R_s^{(l)}(t)$ exhibits a regular oscillation.
Each $i$th ($i=1,2,3,...$) global cycle of $R_s^{(l)}(t)$ begins from its left minimum, passes the central maximum, and ends at the right minimum [also, corresponding to the beginning point of the next $(i+1)$th global cycle]; the 1st global cycle of $R_s^{(l)}(t)$ appears after transient times of $10^3$ ms. Spikes which appear in the $i$th global cycle of $R_s^{(l)}(t)$ forms the $i$th stripe/band in the raster plot. To measure the degree of population synchronization in each of the stimulated ($l=1$) and the non-stimulated ($l=2$) sub-populations, a statistical-mechanical measure $M_s^{(l)}$, based on $R_s^{(l)}(t)$, was introduced by considering the occupation pattern and the pacing pattern of spikes in the stripes/bands \cite{Kim1}. The spiking measures $M_i^{(l)}$ of the $i$th stripe/band [appearing in the $i$th global cycle of $R_s^{(l)}(t)$] is defined by the product of the occupation degree $O_i^{(l)}$ of spikes (representing the density of the $i$th stripe/band) and the  pacing degree $P_i^{(l)}$ of spikes (denoting the smearing of the $i$th stripe/band):
\begin{equation}
  M_i^{(l)} = O_i^{(l)} \cdot P_i^{(l)}.
\label{eq:SM}
\end{equation}
The occupation degrees $O_i^{(l)}$ in the $i$th stripe/band is given by the fractions of spiking neurons in the $i$th stripe/band:
\begin{equation}
   O_i^{(l)} = \frac {N_{i}^{(l)}} {N},
\label{eq:OD}
\end{equation}
where $N_{i}^{(l)}$ is the number of spiking neurons in the $i$th stripe/band.
For full synchronization with fully-occupied stripes/bands, $O_i^{(l)} = 1$, while for sparse synchronization with partially-occupied stripes/bands, $O_i^{(l)} < 1$.
The pacing degree $P_i^{(l)}$ of spikes in the $i$th stripe/band can be determined in a statistical-mechanical way by taking into account their contributions to the macroscopic ISPSR kernel estimate $R_s^{(l)}(t)$.
An instantaneous global phase $\Phi^{(l)}(t)$ of $R_s^{(l)}(t)$ was introduced via linear interpolation in the two successive subregions forming global cycles \cite{Kim1}. The global phase $\Phi^{(l)}(t)$  between the left minimum (corresponding to the beginning point of the $i$th global cycle) and the central maximum is given by
\begin{equation}
\Phi^{(l)}(t) = 2\pi(i-3/2) + \pi \left(\frac{t-t_{i}^{(l,min)}}{t_{i}^{(l,max)}-t_{i}^{(l,min)}} \right) {\rm~~ for~} ~t_{i}^{(l,min)} \leq  t < t_{i}^{(l,max)},
\end{equation}
and $\Phi^{(l)}(t)$ between the central maximum and the right minimum [corresponding to the beginning point of the $(i+1)$th global cycle]
is given by
\begin{equation}
\Phi^{(l)}(t) = 2\pi(i-1) + \pi \left(\frac{t-t_{i}^{(l,max)}}{t_{i+1}^{(l,min)}-t_{i}^{(l,max)}} \right) {\rm~~ for~} ~t_{i}^{(l,max)} \leq  t < t_{i+1}^{(l,min)},
\end{equation}
where $t_{i}^{(l,min)}$ is the beginning time of the $i$th ($i=1, 2, 3, \cdots$) global cycle of $R_s^{(l)}(t)$ [i.e., the time at which the left minimum of $R_s^{(l)}(t)$ appears in the $i$th global cycle], and $t_{i}^{(l,max)}$ is the time at which the maximum of $R_s^{(l)}(t)$ appears in the $i$th global cycle. Then, the contributions of the $k$th microscopic spikes in the $i$th stripe/band occurring at the times $t_{k}^{(l)}$ to $R_s^{(l)}(t)$ is given by $\cos \Phi_k^{(l)}$, where $\Phi_k^{(l)}$ are the global phases at the $k$th spiking time [i.e., $\Phi_k^{(l)} \equiv \Phi^{(l)}(t_{k}^{(l)})$]. Microscopic spikes make the most constructive (in-phase) contributions to $R_s^{(l)}(t)$ when the corresponding global phases $\Phi_k^{(l)}$  is $2 \pi n$ ($n=0,1,2, \dots$), while they make the most destructive (anti-phase) contribution to $R_s^{(l)}(t)$ when $\Phi_k^{(l)}$ is $2 \pi (n-1/2)$. By averaging the contributions of all microscopic spikes in the $i$th stripe/band to $R_s^{(l)}(t)$, we obtain the pacing degrees $P_i^{(l)}$ of spikes in the $i$th stripe/band:
\begin{equation}
P_i^{(l)} = { \frac {1} {S_i^{(l)}}} \sum_{k=1}^{S_i^{(l)}} \cos \Phi_k^{(l)}
\label{eq:PD}
\end{equation}
where $S_i^{(l)}$ is the total number of microscopic spikes in the $i$th stripe/band. By averaging $M_i^{(l)}$ of Eq.~(\ref{eq:SM}) over a sufficiently large number $N_s^{(l)}$ of stripes/bands, we obtain the statistical-mechanical spiking measure $M_s^{(l)}$:
\begin{equation}
M_s^{(l)} =  {\frac {1} {N_s^{(l)}}} \sum_{i=1}^{N_s^{(l)}} M_i^{(l)}.
\label{eq:SM2}
\end{equation}
Here, we follow $3 \times 10^3$ global cycles in each realization, and obtain the average occupation degree, the average pacing degree, and the average statistical-mechanical spiking measure via average over 30 realizations.

\newpage
\begin{table}
\caption{Parameter values used in our computations; units of the capacitance, the potential, the current, the time, and the angular frequency are pF, mV, pA, ms, and rad/ms respectively.}
\label{tab:Parm}
\begin{ruledtabular}
\begin{tabular}{llllll}
(1) & \multicolumn{5}{l}{Izhikevich RS Pyramidal Neurons \cite{Izhi3}} \\
& $C=100$ & $v_r=-60$ & $v_t=-40$ & $v_p=35$ & $v_b=-60$  \\
& $k=0.7$ & $a=0.03$ & $b=-2$ & $c=-50$ & $d=100$  \\
\hline
(2) & \multicolumn{5}{l}{Izhikevich FS Interneurons \cite{Izhi3}} \\
& $C=20$ & $v_r=-55$ & $v_t=-40$ & $v_p=25$ & $v_b=-55$  \\
& $k=1$ & $a=0.2$ & $b=0.025$ & $c=-45$ & $d=0$  \\
\hline
(3) & \multicolumn{5}{l}{External Common Stimulus to Izhikevich Neurons} \\
& $I_{DC} = 70$ & \multicolumn{4}{l}{$D=1$ (RS pyramidal neuron)} \\
& $I_{DC} = 1500$ & \multicolumn{4}{l}{$D=50$ (FS interneuron)} \\
\hline
(4) & \multicolumn{5}{l}{Excitatory AMPA and Inhibitory GABAergic Synapses \cite{Synapse}} \\
& $\tau_l=1$ & $\tau_r=0.5$ & $\tau_d=2$ & \multicolumn{2}{l}{$V_{syn}=0$ (excitatory AMPA synapse)} \\
& $\tau_l=1$ & $\tau_r=0.5$ & $\tau_d=5$ & \multicolumn{2}{l}{$V_{syn}=-80$ (inhibitory GABAergic synapse)} \\
\hline
(5) & \multicolumn{5}{l}{Synaptic Connections between Neurons} \\
& \multicolumn{5}{l}{$M_{syn}=50$ and $p=0.2$ (Watts-Strogatz SWN)} \\
& \multicolumn{5}{l}{$J=15$ (RS pyramidal neurons)} \\
& \multicolumn{5}{l}{$J=100$ and 1000 (FS interneurons)} \\
\hline
(6) & \multicolumn{5}{l}{External Time-Periodic Stimulus to Izhikevich Neurons} \\
& \multicolumn{5}{l}{$A:$ Varying} \\
& \multicolumn{5}{l}{$\omega_d=0.048$ (RS pyramidal neurons)} \\
& \multicolumn{5}{l}{$\omega_d=1.26$ and 0.48 (FS interneurons)}
\end{tabular}
\end{ruledtabular}
\end{table}

\newpage

\newpage

\newpage

\newpage

\newpage

\newpage

\newpage

\newpage

\end{document}